\documentclass[a4paper,11pt]{article}
\pdfoutput=1

\usepackage{jcappub}

\usepackage[T1]{fontenc}

\title{Cosmological perturbations during the Bose-Einstein condensation of dark matter}

\author{R. C. Freitas}
\author{and S. V. B. Gon\c{c}alves}
\affiliation{Universidade Federal do Esp\' \i rito Santo, Centro de Ci\^encias Exatas, Departamento de F\' \i sica, \\Av. Fernando Ferrari, 514 Campus de Goiabeiras, CEP 29075-910, Vit\'oria, Esp\' \i rito Santo, Brazil}

\emailAdd{rodolfo.camargo@pq.cnpq.br}
\emailAdd{sergio.vitorino@pq.cnpq.br}

\abstract{In the present work, we analyze the evolution of the scalar and tensorial perturbations and the quantities relevant for the physical description of the Universe, as the density contrast of the scalar perturbations and the gravitational waves energy density during the Bose-Einstein condensation of dark matter. The behavior of these parameters during the Bose-Einstein phase transition of dark matter is analyzed in details. To study the cosmological dynamics and evolution of scalar and tensorial perturbations in a Universe with and without cosmological constant we use both analytical and numerical methods. The Bose-Einstein phase transition modifies the evolution of gravitational waves of cosmological origin, as well as the process of large-scale structure formation.}


\arxivnumber{1211.6701}

\begin{document}
\maketitle
\flushbottom

\section{Introduction}
\label{sec:intro}

In what concerns the standard cosmological cenarium ($\Lambda$CDM model) the observational data \cite{wmap1, wmap2} show that the ordinary baryonic matter is responsible for approximately only $5~\%$ (neglecting radiation) of the total energy density of the Universe. Nearly $70\%$ of the total density is the contribution of the so called dark energy (DE), responsible for the observed acceleration of the cosmic expansion \cite{riess, perlmutter}. The other $25\%$ is in the form of a weakly interacting matter, called cold dark matter (CDM) or simply dark matter (DM), whose properties are inferred from its gravitational interaction with visible matter, radiation and its effect on the large scale structure of the Universe.

The existence of DM was initially observed  due to discrepancies between the calculated mass of the observed matter and the mass determined from the gravitational effects. Subsequently the observation of the rotational curves of galaxies, the velocity dispersions of galaxies, the gravitational lenses and the large scale structures \cite{dm.evid1, dm.evid2} suggested the existence of the DM. There are several hypothetical candidates for DM particles e.g. weakly interacting massive particles (WIMPs), axions, hidden dark matter, sterile neutrinos and light gravitinos \cite{dm.part} and although the nature of DM is still unkown a great effort has been made in an attempt to detect the DM particles \cite{dm.evid1, dm.evid2, dm.part}.

The BEC process, that is a very well observed phenomenon in terrestrial experiments, occurs when a gas of bosons is cooled at very low temperatures, near absolute zero, what makes a large fraction of the particles occupy the same ground state. This kind of condensation could have occured at some moment during the cosmic history of the Universe. The cosmic BEC mechanism was broadly discussed in \cite{bec1, bec2}. In general the BEC takes place when the gas temperature is below the critical temperature $T_{\textrm{crt}}<2\pi\hbar^{2}n^{2/3}/mk_{B}$, where $n$ is the particles density, $m$ is the particle mass and $k_B$ is the Boltzmann's constant. Since in an adiabatic process a matter dominated Universe behaves as $\rho\propto T^{3/2}$ the cosmic dynamics has the same temperature dependence. Hence we will have the critical temperature at present $T_{\textrm{crt}}=0.0027~\textrm{K}$ if the boson temperature was equal to the radiation temperature at the red-shift $z=1000$ \cite{bec2}. During the cosmic adiabatic evolution the ratio of the photon temperature and the matter temperature evolves as $T_{\textrm{r}}/T_{\textrm{m}}~\propto~a$, where $a$ is the scale factor of the Universe. Using as value for the present energy density of the Universe $\rho=9.44\times10^{-30} \textrm{g}/\textrm{cm}^{3}$ BEC will happens if the boson mass satisfies $m<~1.87\textrm{eV}$.

Despite the great sucess and simplicity of the standard $\Lambda$CDM cosmological model, it still remains as a phenomenological model and we can not explain why the cosmological constant density has a value close to the critical density today and why the measured cosmological constant is smaller than the quantum vacuum energy density by a factor of $10^{-120}$.  At galactic scales, this model has difficulties in explaining the observed distribution of DM around the luminous one. The simulations in the $\Lambda$CDM model predict that bound halos around galaxies must have very characteristic density profiles with a well pronounced central cusp \cite{cuspcore}, but the observations of rotation curves show that the distribution of DM is shallower than the predicted and it has a approximately constant density core \cite{rotcurve}. The so called core-cusp problem can be solved by presuming that DM particles are ultralight scalar particles with masses of the order of $10^{-22}~\textrm{eV}$ initially in a cold BEC \cite{fucdm}.

Assuming the hypothesis that the CDM in a galaxy is in form of BEC the density distribution of the static gravitationally bounded single component BEC DM is given by $\rho(r)=\rho_{\ast}\sin{kr}/kr$, where $\rho_{\ast}=\rho(0)$ is the density in the center of the condensate and $k$ is a constant. Giving the conditions $\rho(R)=0$ and $kR=\pi$, where $R$ is the condensate radius, the condensate DM halo radius can be fixed as
$R=\pi(\hbar^{2}l_{s}/ G m^{3})^{1/2}$ \cite{harko1} where $l_s$ is the particle scattering length and $m$ is the the particle mass. The calculated total mass of the condensate DM halo is $M=4\pi^{2}(\hbar^{2}l_s/Gm^{3})^{3/2}\rho_{\ast}=4R^{3}\rho_{\ast}/\pi$ \cite{harko1}. So the mass of the particles of the condensate is \cite{harko1}
\begin{equation}
   m=\left(\frac{\pi^2\hbar^2l_s}{GR^2}\right)^{1/3} \approx 6.73\times10^{-2}\left(\frac{l_s}{1~\textrm{fm}}\right)^{1/2}\left(\frac{R}{1~\textrm{kpc}}\right)^{-2/3}~\textrm{eV} \quad.
\end{equation}
\par Recently the cosmological BEC process of DM was investigated and in this model \cite{harko2} it is assumed that the condensation process is a phase transition that occurs at some time during the history of the Universe. In this model the normal bosonic DM cools below the critical condensation temperature, that turns to be favorable to form the condensate in which all particles occupy the same ground state. In this new period the two phases coexist for some time until all ordinary DM is converted into the condensated form, when the transition ends. The time evolution of cosmological parameters as the energy density, temperature and scale factor is changed during the phase transition process \cite{harko2}.

In this letter we aim to investigate how the generation of cosmological perturbations produced by the space-time expansion, and the parameter associated with them, would be changed during this Bose-Einstein phase transition of dark matter. The scalar perturbations are related to the formation of structures in the Universe, like galaxy clusters and galaxy superclusters. The generally accepted picture of the structure formation in cosmology is the one where the Universe started off in an extremely homogeneous and isotropic state, with initial conditions provided by an era of accelerated expansion called inflation. The tiny primordial density fluctuations, generated during inflation from quantum fluctuations of the vacuum, would later grow under the influence of gravity and eventually collapse to form the structures that we observe today. In general, the evolution of the density of the different field components of the matter are calculated analytically or numerically depending on the complexity of the studied model. These calculations are made in such a way that they may be compared with the observations of large-scale structure of the Universe like, e.g., Cosmic Microwave Background (CMB) anisotropy maps derived from WMAP data \cite{wmap1, wmap2}, Type Ia supernova (SNIa) surveys \cite{riess, perlmutter} and the baryon acoustic oscillations (BAO) in the Sloan Digital Sky Survey (SDSS) luminous galaxy sample \cite{BAO}.

The gravitational waves are tensorial fluctuations in the metric of space-time. This particular perturbation is not explicitly coupled with the energy density and pressure of the matter of the Universe and does not contribute to the gravitational instability that generates the cosmological structures we see today. On the other hand this study is of great interest because it supplies the specific signature of the metric theory of gravity. These waves are predicted by Einstein's theory of General Relativity (GR) but they still have to be directly detected. Great efforts have been done in this sense and there is hope that a new generation of experiments in space may allow this detection \cite{gwobs}. The gravitational waves spectrum frequency extends over a wide range of interest, from $10^{-18}Hz$ to $10^{8}Hz$, depending on the sources that generate those waves. 

The present letter is organized as follows. The basic properties of the Bose-Einstein condensate dark matter (BEC DM) and the ordinary dark matter are reviewed in Section \ref{sec:dm}. The cosmological dynamics of the density perturbations is considered in Section \ref{sec:scalarperturbation}. The evolution of the gravitational waves is derived in Section \ref{sec:gw}. We discuss and conclude our results in Section \ref{sec:conc}.

\section{Ordinary Dark Matter and Bose-Einstein condensate Dark Matter}
\label{sec:dm}

Following the present data \cite{wmap1,wmap2} we assume a flat homogeneous and isotropic Universe, whose geometry is described by the Friedmann-Robertson-Walker metric, given by
\begin{equation}
   ds^2=c^2dt^2-a^2(t)d\vec{x}^2 \quad ,
\end{equation}
where $a(t)$ is the scale factor of the Universe that describes the cosmic evolution, $t$ is the cosmic time and $c$ is the speed of light. The gravitational dynamics is given by the Einstein's field equations
\begin{equation}
   \label{eq.einstein}
   R_{\mu\nu}-\frac{1}{2}g_{\mu\nu}R=\frac{8\pi G}{c^4}T_{\mu\nu} \quad.
\end{equation}
We also consider the Universe filled by a perfect fluid, described by the energy-momentum tensor
\begin{equation}
   T^{\mu\nu}=(\rho c^2+p)u^\mu u^\nu - pg^{\mu\nu} \quad ,
\end{equation}
where $\rho$ is the density of the fluid, $p$ is the pressure and $g^{\mu\nu}$ is the metric tensor.

Imagining the Universe filled by baryonic matter, radiation, dark matter and cosmological constant and neglecting any possible interaction between those components by assuming that the energy of each component is individually conserved, the equations of motion are 
\begin{eqnarray}
   \label{eq.friedmann}
   \frac{\dot{a}^2}{a^2} & = & \frac{8\pi G}{3}(\rho_b + \rho_r + \rho_\chi+\rho_\Lambda) \quad , \\
   2\frac{\ddot{a}}{a} & + & \frac{\dot{a}^2}{a^2} =-\frac{8\pi G}{c^2}(p_r+p_\chi+p_\Lambda) \quad , \\
   \label{eq.conseng}
   \dot{\rho_i} & + & 3\left(\rho_i+\frac{p_i}{c^2}\right)\frac{\dot{a}}{a}  =  0 \quad .
\end{eqnarray}
where $H=\dot{a}/a$ is the Hubble parameter and $i=b,~r,~\chi,~\Lambda~$ means baryonic matter, radiation, DM and comological constant, respectively.

\subsection{Non-condensated Dark Matter}

We will presume that in the early phases of the Universe the DM was made of bosons particles with mass $m_\chi$ and temperature $T$, originated in equilibrium and decoupled at a temperature $T_D$. Particles that decouple from the plasma in the primordial Universe at any temperature $T_D$ has its momenta redshifted such that the distribution function at any time after the decoupling is related to the value of the distribution function at the moment of the decoupling and it keeps an equilibrium shape in both non-relativistic and extreme-relativistic regimes \cite{harko2}. In the non-relativistic regime the non-condensated DM pressure is \cite{harko2}
\begin{equation}
   \label{eq.eos1}
   p_\chi=c^2\sigma^2\rho_\chi \quad ,
\end{equation}
where $\sigma^2=\left\langle \vec{v}^{~2}\right\rangle/3c^2$ is the velocity dispersion and $\left\langle \vec{v}^{~2}\right\rangle$ is the mean square velocity of the particle. Applying the equation of state (\ref{eq.eos1}) into the conservation equation (\ref{eq.conseng}) leads to
\begin{equation}
   \label{eq.normDM}
   \rho_\chi=\frac{\rho_\chi^{(0)}}{(a/a_0)^{3(1+\sigma^2)}} \quad ,
\end{equation}
where $\rho=\rho_\chi^{(0)}$ when $a=a_0$. Replacing the evolution equation (\ref{eq.normDM}) for the non-condensated DM into the  Friedmann's equation (\ref{eq.friedmann}) we find that the evolution of a Universe filled by baryonic matter, radiation, non-condensated DM and cosmological constant is given by
\begin{equation} \frac{\dot{a}^2}{a^2}=H_0^2\left(\frac{\Omega_b^{(0)}}{(a/a_0)^3}+\frac{\Omega_r^{(0)}}{(a/a_0)^4}+\frac{\Omega_\chi^{(0)}}{(a/a_0)^{3(1+\sigma^2)}}+\Omega_\Lambda\right) \quad ,
\end{equation}
where $H_0$ is the Hubble parameter and $\Omega_i^{0}=8\pi G\rho_i^{(0)}/3H_0^2$ is the density parameter, both at $a=a_0$. The density parameters satisfy the relation $\Omega_b^{(0)}+\Omega_r^{(0)}+\Omega_\chi^{(0)}+\Omega_\Lambda=1$ and $a_0$ can be considered the present day scale factor. Since the DM is non-relativistic the cosmic evolution is little dominated by the term $\sigma^2$. In the standard model $\sigma^2=0$.

\subsection{Bose-Einstein condensate Dark Matter}

The BEC is formed when all particles in a Bose gas occupy the same quantum ground state. This happens at a very low critical temperature $T_{\textrm{crt}} \approx 2\pi \hbar^2\rho^{2/3}/m^{5/3}k_B$ \cite{bec1, bec2, harko2}, where $m$ is the mass of the particles, $\rho$ is the density and $k_B$ is the Boltzmann's constant. The coherent state is reached when the temperature is low enough or the density is sufficiently high \cite{harko2}. In this model we will assume that almost all the DM particles are in the condensate form and that the DM halos are made of BEC DM at absolute zero temperature.

Only collisions between two particles at low energy are relevants in a diluate and cold gas. These collisions are characterized by the scattering length $l_s$ independently of the two-body potential \cite{harko2}, hence the interaction potencial can be replaced by an effective interaction $V_I(\vec{r}~'-\vec{r})=\lambda\delta(\vec{r}~'-\vec{r})$ \cite{bec3, harko2} where $\lambda=4\pi\hbar^2l_s/m$. The ground state features of BEC DM halos is described by the generalized Gross-Pitaevskii (GP) equation \cite{bec3, harko2}
\begin{equation}
   \label{eq.gGP}
   \dot{\imath}\hbar\frac{\partial \phi(t,\vec{r})}{\partial t}=-\frac{\hbar^2}{2m}\nabla^2\phi(t,\vec{r})+mV(\vec{r})\phi(t,\vec{r})+U_0|\phi(t,\vec{r})|^2\phi(t,\vec{r}) \quad ,
\end{equation}
where $\phi(t,\vec{r})$ is the wave function of the condensate, $m$ is the particles mass, $V$ is tha gravitational potential that satisfies the Poisson's equation $\nabla^2V(\vec{r})=4\pi G\rho$ and $U_0=4\pi\hbar^2l_s/m$. We can use the Madelung representation of the wave function \cite{bec1, bec2, bec3}, which is
\begin{equation}
   \phi(t,\vec{r})=\sqrt{\rho(t,\vec{r})}\times e^{\dot{\imath}S(t,\vec{r})/\hbar} \quad ,
\end{equation}
where $\rho(t,\vec{r})=|\phi(t,\vec{r})|^2$ is the density of the condensate and the function $S(t,\vec{r})$ has the dimension of an action. This transformation will make the generalized GP equation (\ref{eq.gGP}) breaks into two equations
\begin{eqnarray}
   \frac{\partial \rho_\chi}{\partial t} +\nabla \cdot(\rho_\chi\vec{v}) & = & 0 \quad , \\
   \rho_\chi\left(\frac{\partial \vec{v}}{\partial t}+(\vec{v}\cdot \nabla)\vec{v}\right) & = & -\nabla p_\chi\left(\frac{\rho_\chi}{m}\right)-\rho_\chi\nabla\left(\frac{V}{m}\right) -\nabla V_Q \quad ,
\end{eqnarray}
where $V_Q=-(\hbar^2/2m)\nabla^2\sqrt{\rho_\chi}/\rho_\chi$ is a quantum potential and $\vec{v}=\nabla S/m$ is the velocity of the quantum fluid. The effective pressure of the condensate \cite{bec1, bec2, harko2} is given by
\begin{equation}
   \label{eq.eos2}
   p_\chi\left(\frac{\rho_\chi}{m}\right)=\frac{2\pi \hbar^2l_s}{m^3} \rho_\chi^2 \quad ,
\end{equation}
and the dynamics of the fluid is determined by the scatering lenth $l_s$ and the mass $m$ of the DM particles. Using the conservation equation (\ref{eq.conseng}) and the equation of state (\ref{eq.eos2}) we find
\begin{equation}
   \dot{\rho_\chi}+3\rho_\chi \left(1+\omega_{\chi}\rho_\chi\right)\frac{\dot{a}}{a}=0 \quad ,
\end{equation}
with $\omega_{\chi}=2\pi \hbar^2l_s/m^3c^2$. The evolution of the BEC DM density is given by
\begin{equation}
   \rho_\chi = \frac{C}{(a/a_0)^3-\omega_{\chi}C} \quad ,
\end{equation}
where $C$ is an integration constant. Ensuring that $\rho_\chi=\rho_\chi^{(0)}$ at $a=a_0$ we will obtain
\begin{equation}
   \rho_\chi=\frac{1}{\omega_{\chi}}\frac{\rho_{0\chi}}{(a/a_0)^3-\rho_{0\chi}} \quad ,
\end{equation}
with 
\begin{equation}
   \rho_{0\chi}=\frac{\omega_{\chi}\rho_\chi^{(0)}}{1+\omega_{\chi}\rho_\chi^{(0)}} = \frac{\omega_{\chi}\rho_{crt}^{(0)}\Omega_\chi^{(0)}}{1+\omega_{\chi}\rho_{crt}^{(0)}\Omega_\chi^{(0)}} \quad ,
\end{equation}
where the critical density of the Universe today is $\rho_{crt}^{(0)}=9.24 \times 10^{-27}~\textrm{Kg}/\textrm{m}^3$. And the Friedmann's equation (\ref{eq.friedmann}) for the BEC DM is 
\begin{equation}
\label{eq.BECDMsclfact}
\frac{\dot{a}^2}{a^2}=H_0^2\left(\frac{\Omega_b^{(0)}}{(a/a_0)^3}+\frac{\Omega_r^{(0)}}{(a/a_0)^4}+\frac{1}{\omega_\chi}\frac{\Omega_{0\chi}}{(a/a_0)^3-\rho_{0\chi}}+\Omega_\Lambda\right) \quad ,
\end{equation}
where $\Omega_{0\chi}=\rho_{0\chi}/\rho_{crt}^{(0)}$.

\subsection{Cosmological dynamics during the condensation phase}

The best description of the BEC dynamics seems to be a first order phase transition \cite{harko2}, since it was demonstrated, by analyzing the temperature  and density dependence of the chemical potential in several theoretical models that describe the thermodynamical transition from the normal phase to the BEC phase, that none of them predicts a second-order phase transition \cite{harko2, phasetrans1, phasetrans2}. On the other hand a first order phase transition characterize a genuine mathematical singularity, but for the case of a ideal Bose gas confined in a cubic box it was shown that a system with a finite number of particles can present a discontinuous phase transition that characterize a genuine mathematical singularity, provided that the pressure is kept constant \cite{phasetrans2}. 

To see it more clearly let us consider some thermodynamical aspects of the process. Due to the extensivity property of the Helmholtz free energy $F(N,V,T)$, where $N$ is the total particles number of the system, $V$ is the total volume and $T$ is the temperature, we can write
\begin{eqnarray}
   f(n,T) & = & \frac{F}{V} \quad , \\
	\tilde{f}(v,T) & = & \frac{F}{N} \quad ,
\end{eqnarray}
with $v=N/V=1/n$. It is easy to show that
\begin{equation}
   f(n,T)=n\tilde{f}(v,T) \quad .
\end{equation}
\par The chemical potential $\mu=\mu(n,T)$ of this physical system is a function of the particle density $n$ and the temperature $T$, and from the Helmholtz free energy we can find that
\begin{eqnarray}
    \mu(n,T) & = & \left(\frac{\partial F}{\partial N}\right)_{T,V}=\left(\frac{\partial f(n,T)}{\partial n}\right)_{T,V} \quad , \\
		p(v,T) & = & -\left(\frac{\partial F}{\partial V}\right)_{T,N}=-\left(\frac{\partial \tilde{f}(v,T)}{\partial v}\right)_{T,N} \quad . 
\end{eqnarray}
From the relations above we can see that the pressure $p(v,T)$ and the chemical potential $\mu(n,T)$ carry the same information about the physical system. The chemical potential $\mu(n,T)$ contains fundamental information about the nature of the phase transition, which can be inferred from the isothermal compressibility
\begin{equation}
   \kappa_{T}^{-1}=n^2\left(\frac{\partial \mu}{\partial n}\right)_{T} = - v\left(\frac{\partial p}{\partial v}\right)_{T} \quad .
\end{equation}
\par The laws of thermodynamics require that both the chemical potential and the pressure have only one value for each pair of fixed $(n,T)$ and $(v,T)$ respectively. That is, they must be single valued. When $(\partial p/\partial v)_{T}<0$ the system is stable. For $(\partial p/\partial v)_{T}\geq 0$ we have a first order phase transition and the volume at the phase transition is not single valued and for $(\partial p/\partial v)_{T}=0$ a genuine singularity develops and we will have a discontinuous phase transition.

\par Therefore, during the quasi-static process of the BEC phase transition, the thermodynamic condition that has to be satisfied is the continuity of the pressure at the transition point, which restrict the BEC parameters. So the equations of state (\ref{eq.eos1}) and (\ref{eq.eos2}) give the critical transition density \cite{harko2}
\begin{equation}
   \rho_{\chi}^{crt}=\frac{c^2\sigma^2m^3}{2\pi\hbar^2l_s}=\frac{\sigma^2}{\omega_\chi} \quad ,
\end{equation}
the critical temperature \cite{harko2, bec3} is
\begin{equation}
   T_{crt}=\frac{2\pi\hbar^2}{\zeta(3/2)^{2/3}m^{5/3}k_B}\left(\rho_\chi^{crt}\right)^{2/3} \quad ,
\end{equation} 
where $\zeta(3/2)$ is the Riemann zeta function. The scale factor of the Universe at the moment of the condensation is \cite{harko2}
\begin{equation}
   \frac{a_{crt}}{a_0}=\left(\frac{2\pi\hbar^2l_s}{c^2\sigma^2m^3}\rho_{crt}^{(0)}\Omega_{\chi}^{(0)}\right)^{\frac{1}{3(1+\sigma^2)}} \quad ,
\end{equation}
and the critical redshift is 
\begin{equation}
   \label{eq:zcrit}
   1+z_{crt}=\left(\frac{2\pi\hbar^2l_s}{c^2\sigma^2m^3}\rho_{crt}^{(0)}\Omega_{\chi}^{(0)}\right)^{\frac{-1}{3(1+\sigma^2)}} \quad .
\end{equation}

During the transition phase the pressure $P=P_{crt}$ and the temperature $T=T_{crt}$ are constant and after the beginning of the phase transition the DM density decreases from $\rho_\chi^{norm}$, when all DM is in non-condensated form, to $\rho_\chi^{BEC}$ when all DM is condensated, when the transition phase ends. So we introduce the volume fraction of matter in the BEC phase \cite{harko2}
\begin{equation}
   f(t)=\frac{\rho_\chi(t)-\rho_\chi^{norm}}{\rho_\chi^{BEC}-\rho_\chi^{norm}} \quad ,
\end{equation}
and we can write
\begin{eqnarray}
   \rho_\chi(t) & = & \rho_\chi^{norm}\left(1+n_\chi f(t)\right) \quad , \\
   n_\chi & = & \frac{\rho_\chi^{BEC}-\rho_\chi^{norm}}{\rho_\chi^{norm}} \quad .
\end{eqnarray}

When the condensation starts, at $t=t_{crt}$ all the DM is in non-condensated form, i.e., $\rho_\chi(t_{crt})=\rho_\chi^{norm}$ and $f(t_{crt})=0$. At the end of the condensation all the DM is in BEC form, $\rho_\chi(t_{BEC})=\rho_\chi^{BEC}$, and $f(t_{BEC})=1$. With help of the energy conservation equation (\ref{eq.conseng}) we find that \cite{harko2}
\begin{eqnarray}
   \label{eq.sclfactor}
   a(t) & = & a_{crt}\left(1+rf(t)\right)^{-1/3} \quad , \\
   r & = & \frac{n_\chi}{1+P_{crt}/\rho_\chi^{norm}c^2} \quad ,
\end{eqnarray}
where $a(t_{crt})=a_{crt}$ and the phase transition ends at
\begin{equation}
   1+z_{BEC}=(1+r)^{1/3}(1+z_{crt}) \quad .
\end{equation}

Using the Friedmann's equation (\ref{eq.friedmann}) we find that the evolution of the volume fraction is given by
\begin{equation}
\label{eq.difmassfrac} \frac{df}{d\tau}=-3\left(\frac{1+rf}{r}\right)\sqrt{\frac{\Omega_b^{(0)}}{(a_{crt}/a_0)^3}(1+rf)+\frac{\Omega_r^{(0)}}{(a_{crt}/a_0)^4}(1+rf)^{4/3}+\Omega_\chi^{(0)}(1+n_\chi f)+\Omega_\Lambda} \quad ,
\end{equation}
where $\Omega_\chi^{(0)}=\rho_\chi^{norm}/\rho_{crt}^{(0)}$ and $\tau=H_0t$. Since $\rho_\chi^{BEC}<\rho_\chi^{norm}$ we will have $r\in(-1,0)$ and $n_\chi<0$. On the other hand $P_{crt} / \rho_\chi^{norm}c^2=\sigma^2<<1$ and from equation (\ref{eq.sclfactor}) we can approximate $r\approx n_\chi$ \cite{harko2} and if we neglect the radiation contribution to the energy density the equation (\ref{eq.difmassfrac}) has the solution 
\begin{equation}
   \label{eq.fLCDM}
   f(t)=\frac{\Omega_\Lambda^2}{r\Omega_g}\left[\frac{1+\Omega_c e^{-3(\Omega_\Lambda/H_0^{-1})(t-t_{crt})}}{1-\Omega_c e^{-3(\Omega_\Lambda/H_0^{-1})(t-t_{crt})}}\right]^2-\frac{\Omega_\Lambda+\Omega_g}{r\Omega_g} \quad ,
\end{equation}
where $H_0^{-1}$ is the Hubble time and
\begin{eqnarray}
   \label{eq.Omegag}
   \Omega_g & = & \frac{\Omega_b^{(0)}}{(a_{crt}/a_0)^3}+\Omega_\chi^{(0)} \quad , \\
   \label{eq.Omegac}
   \Omega_c & = & \frac{\sqrt{\Omega_g+\Omega_\Lambda}-\Omega_\Lambda}{\sqrt{\Omega_g+\Omega_\Lambda}+\Omega_\Lambda} \quad .
\end{eqnarray}

If we neglect both radiation and cosmological constant contribution to the total energy density it is straightforward to find
\begin{equation}
   \label{eq.fCDM}
   f(t)=\frac{1}{r}\left(1+\frac{2}{3}H_0\sqrt{\Omega_g}(t-t_{crt})\right)^{-2}-\frac{1}{r} \quad .
\end{equation}

When the condensation process ends, for $t\geq t_{BEC}$, the evolution of the Universe is given by equation (\ref{eq.BECDMsclfact}). As can be seen from the equations (\ref{eq.fLCDM}) and (\ref{eq.fCDM}) the scale factor do not depend on $r$.

\begin{figure}[tbp]
   \centering
   \includegraphics[width=0.7\textwidth]{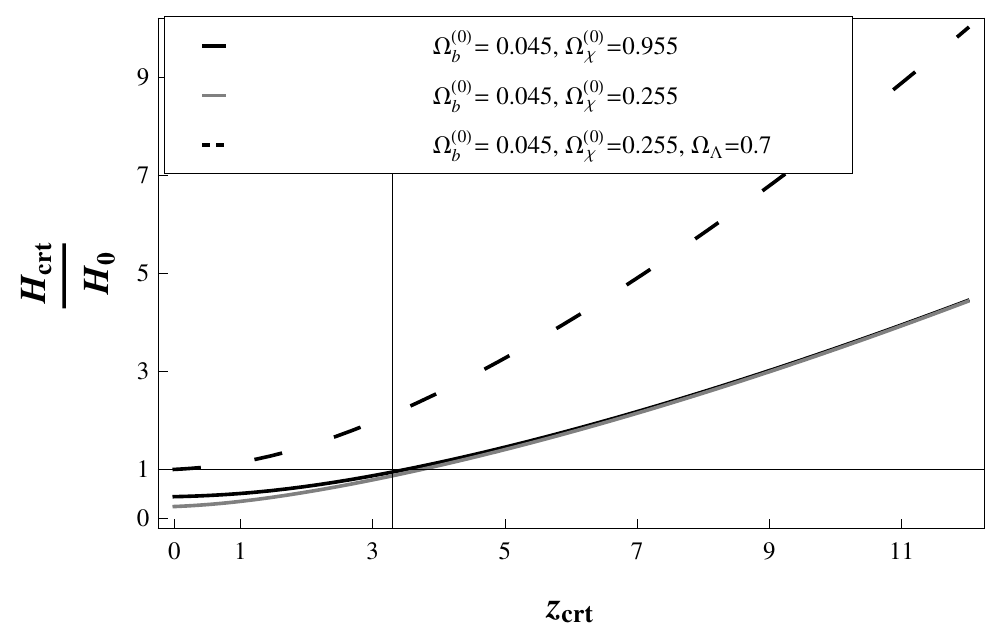}
   \caption{The ratio between the Hubble parameter at the beginning of the condensation and the Hubble parameter today for a Universe filled with cosmological constant (dashed line) and without cosmological constant (solid lines).}
   \label{fig.hubbcrit}
\end{figure}

With equations (\ref{eq.sclfactor}), (\ref{eq.fLCDM}) and (\ref{eq.fCDM}) we calculate the Hubble parameter at the critical point for the model with and without cosmological constant, respectively
\begin{eqnarray}
   \label{eq.hubblepar1}
   H_{crt} & = & \frac{4}{9}H_0\sqrt{\Omega_b^{(0)}(1+z_{crt})^3+\Omega_\chi^{(0)}} \quad , \\
   \label{eq.hubblepar2}
   H_{crt} & = & H_0\sqrt{\Omega_b^{(0)}(1+z_{crt})^3+\Omega_\chi^{(0)}+\Omega_{\Lambda}} \quad ,
\end{eqnarray}
which allow us to constraint the critical redshift $z_{crt}$, since the Hubble parameter at the critical point must be bigger than the one today. In Figure (\ref{fig.hubbcrit}) we see the ratio between the Hubble parameter at the beginning of the condensation process and the Hubble parameter today as a fuction of the critical redshift $z_{crt}$. The requirement that $H_{crt}/H_0>1$ is satisfied for $z_{crt}>3.5$, $z_{crt}>3.75$ and $z_{crt}>0$ respectively.

The effective equation of state parameter $\omega$ is
\begin{equation}
   \label{eq.omega}
   \omega=\frac{p}{\rho c^2}=\frac{\sigma^2\Omega_{\chi}^{(0)}(1+z_{crt})^{-3}-\Omega_{\Lambda}R^3}{\Omega_b^{(0)}+\Omega_{\chi}^{(0)}(1+z_{crt})^{-3}+\Omega_{\Lambda}R^3} \quad ,
\end{equation}
where $R=a/a_0$. This expression and the conservation equation (\ref{eq.conseng}) allow us to calculate the speed of sound $c_s^2=\partial p/\partial \rho$, that is
\begin{equation}
   \label{eq.vsom1}
   \frac{c_s^2}{c^2}=\frac{\sigma^2\Omega_{\chi}^{(0)}(1+z_{crt})^{-3}}{\Omega_b^{(0)}+\Omega_{\chi}^{(0)}(1+z_{crt})^{-3}+\Omega_{\Lambda}R^3}
\quad ,
\end{equation}
which we can approximate to null, since $\sigma^2<<1$.

\section{Cosmological density perturbations}
\label{sec:scalarperturbation}

There are other studies about this subject: with Newtonian gravity \cite{NG}, with the new-Newtonian gravity \cite{NNG, Vel}, with Post-Newtonian approximation by using the conservation of the general relativistic energy-momentum tensor, and considering the small velocity limit \cite{PNA} and the relativistic case \cite{RC01} where the model unifies the dark matter and dark energy.

Here we study the evolution of the structure formation during the BEC phase transition using the gauge-invariant formalism \cite{muka, pert} for the classical linear perturbations. Whe choose the longitudinal (or conformal-Newtonian) gauge, in which for a diagonal energy-momentum tensor the perturbed metric is
\begin{equation}
   \label{eq.pertscalar}
   ds^2=a(\eta)^2\left[(1+2\Phi)d\eta^2-(1-2\Phi)d\vec{x}^2\right] \quad ,
\end{equation}
where $\Phi$ is the gauge-invariant scalar perturbation, which can be interpreted as the relativistic generalization of the Newtonian potential and $\eta$ is the conformal time, related to the cosmic time $t$ by $cdt=ad\eta$. We replace the metric (\ref{eq.pertscalar}) in the Einstein's field equations (\ref{eq.einstein}) and keeping only the first order terms we get the equations of motion for the scalar potential $\Phi$
\begin{eqnarray}
   \label{eq.movpertscalar1}
   \nabla^2\Phi-3\mathcal{H}\left(\Phi'+\mathcal{H}\Phi\right) & = & 4\pi a^2G\delta \rho \quad , \\
   \Phi'' + 3\mathcal{H}\Phi'+\left(2\mathcal{H}'+\mathcal{H}^2\right)\Phi & = & 4\pi G a^2\delta p/c^2 \quad , 
\end{eqnarray}
where the apostrophe $'$ means the derivative to the conformal time $\eta$, $\mathcal{H}=a'/a$ is the conformal Hubble parameter and $\delta \rho$ and $\delta p$ are the gauge-invariant perturbations of the total density and the total pressure, respectively. 

For the case of adiabatic perturbation \cite{muka, pert} we use the relation 
\begin{equation}
   \label{eq.vsom}
   \delta p = c_s^2\delta \rho \quad ,
\end{equation}
combined with the equations (\ref{eq.movpertscalar1}), where $c_s^2=\partial p/\partial \rho$ is the fluid sound speed, to find 
\begin{equation}    
   \Phi''+3\mathcal{H}\left(1+\frac{c_s^2}{c^2}\right)\Phi'+\left[2\mathcal{H}'+\mathcal{H}^2\left(1+3\frac{c_s^2}{c^2}\right)+\frac{c_s^2}{c^2}k^2\right]\Phi=0 \quad ,
\end{equation}
where $k^2$ is the wave-number and we made $\nabla^2\Phi=-k^2\Phi$. With help of a change of variable \cite{pert} and the background equations we find
\begin{equation}
   \label{eq.movpertscalar2}
   a^2\frac{d^2\Phi}{da^2}+\left[4+3\frac{c_s^2}{c^2}-\frac{1}{2}\left(1+3\frac{p}{\rho c^2}\right)\right]a\frac{d\Phi}{da}+\left[1+3\frac{c_s^2}{c^2}\left(1+\frac{k^2}{8\pi G \rho a^2}\right)-\frac{3}{2}\left(1+3\frac{p}{\rho c^2}\right)\right]\Phi=0 \quad .
\end{equation}

To examine the density perturbations $\delta \rho$ we calculate the relations $T^{\mu\nu}_{~;\nu}=0$ up to first order, where $;\nu$ is the covariant derivative, to find \cite{muka, pert}
\begin{eqnarray}
   \label{eq.conspert1a}
   \delta \rho' +\mathcal{H}(\delta \rho +\delta p)-3\Phi'(\rho + p)+a(\rho+p)\delta u^{i}_{~,i} =0 \quad , \\
   \label{eq.conspert1b}
   a^{-4}\left[a^5(\rho +p)\delta u^{i}_{~,i}\right]'+\nabla^2\delta p+(\rho+p)\nabla^2\Phi=0 \quad ,
\end{eqnarray}
where $\delta u^i$ is the fluid velocity perturbation and $,i$ is the ordinary spatial derivative. 
 
In this analysis the baryons and DM do not interact except through gravity and we neglect the radiation contribution. So each component satisfy the pair of equations (\ref{eq.conspert1a}) and (\ref{eq.conspert1b}) separately. Since $P_{crt}/\rho_{\chi}c^2=\sigma^2<<1$ is true we can make an approximation to find that both baryons and DM satisfy the same differential equation
\begin{eqnarray}
   \label{eq.conspert2a}
   (\delta -3\Phi)' & + & a\delta u^{i}_{~,i}=0 \quad , \\
   \delta = \frac{\delta \rho_b}{\rho_b} & = & \frac{\delta \rho_\chi}{\rho_\chi} \quad ,
\end{eqnarray}
where $\delta$ is the density contrast.

Combining equations (\ref{eq.conspert1b}) and (\ref{eq.conspert2a}) and making a change of variables for $R = a/a_0$ we find that
\begin{equation}
   \label{eq.conspert3}
   \frac{d}{dR}\left[R^2\mathcal{H}\frac{d}{dR}\left(\delta-3\Phi\right)\right]-\frac{\nabla^2\Phi}{\mathcal{H}}=0 \quad .
\end{equation}

From equation (\ref{eq.omega}) we see that
\begin{equation}
   \label{eq.omegaaprrox}
   \omega \approx
   \begin{cases}
      \quad \quad 0  \quad \quad , \quad \Omega_{\Lambda}=0 \quad , \\
      \frac{-\Omega_{\Lambda}R^3}{\Omega_{m}+\Omega_{\Lambda}R^3}  \quad , \quad \Omega_{\Lambda}\neq 0 
   \end{cases}
   \quad ,
\end{equation}
where $\Omega_m = \Omega_b^{(0)} + \Omega_{\chi}^{(0)}(1 + z_{crt})^{-3}$. From equation (\ref{eq.vsom1}) for the sound speed we have
\begin{equation}
   \label{eq.vsomaprox}
   \frac{c_s^2}{c^2} \approx 0 \quad ,
\end{equation}
since we consider $\sigma^2<<1$.

\subsection{Universe filled with baryons and DM}

\begin{figure}[tbp]
   \includegraphics[width=0.55\textwidth]{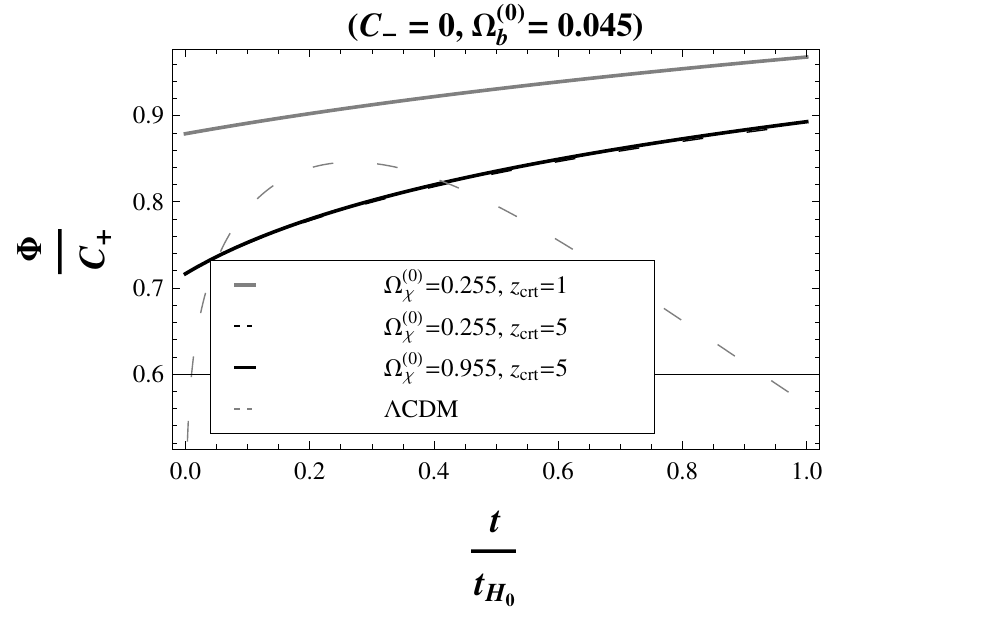}
   \includegraphics[width=0.55\textwidth]{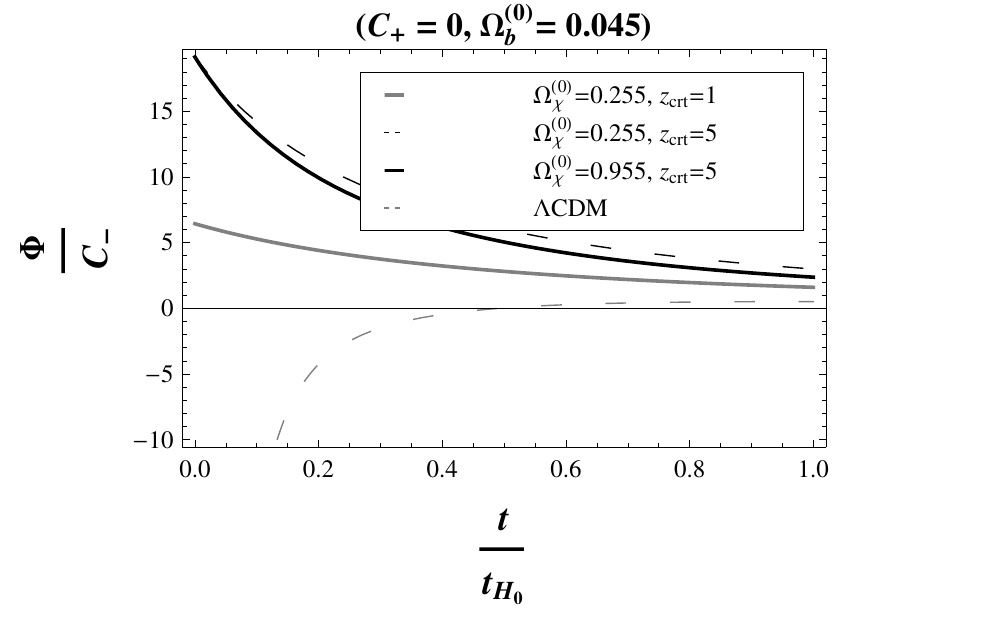}
   \caption{Growing and decaying modes of the potential $\Phi$ as a function of time for the BEC model and $\Lambda$CDM, where in the BEC model the Universe is filled with baryons and DM. In the legends $\Lambda$CDM stands for the concordance model, where $\Omega_b^{0}=0.045$, $\Omega_{dm}^{0}=0.255$ and $\Omega_{\Lambda}=0.7$.}
   \label{fig.PhiCDM}
\end{figure}

With equations (\ref{eq.aproxCDM}) and (\ref{eq.vsomaprox}) the perturbation equation (\ref{eq.movpertscalar2}) will be
\begin{equation}
   R^2\frac{d^2\Phi}{dR^2}+\frac{7}{2}R\frac{d\Phi}{dR}-\frac{1}{2}\Phi=0 \quad ,
\end{equation}
which has a simple power law solution
\begin{eqnarray}
   \label{eq.solescalar1a}
   \Phi & = & C_{+}R^{n_{+}}+C_{-}R^{n_{-}} \quad , \\
   n_{\pm} & = & \frac{-5\pm\sqrt{33}}{4} \quad ,
\end{eqnarray}
and the growing and decaying modes of the potential $\Phi$ as a function of time for the BEC model with several different values for the model parameters and also for the concordance model ($\Lambda$CDM, with $\Omega_b^{0}=0.045$, $\Omega_{dm}^{0}=0.255$ and $\Omega_{\Lambda}=0.7$) are shown in Figure (\ref{fig.PhiCDM}).

\begin{figure}[tbp]
   \includegraphics[width=0.55\textwidth]{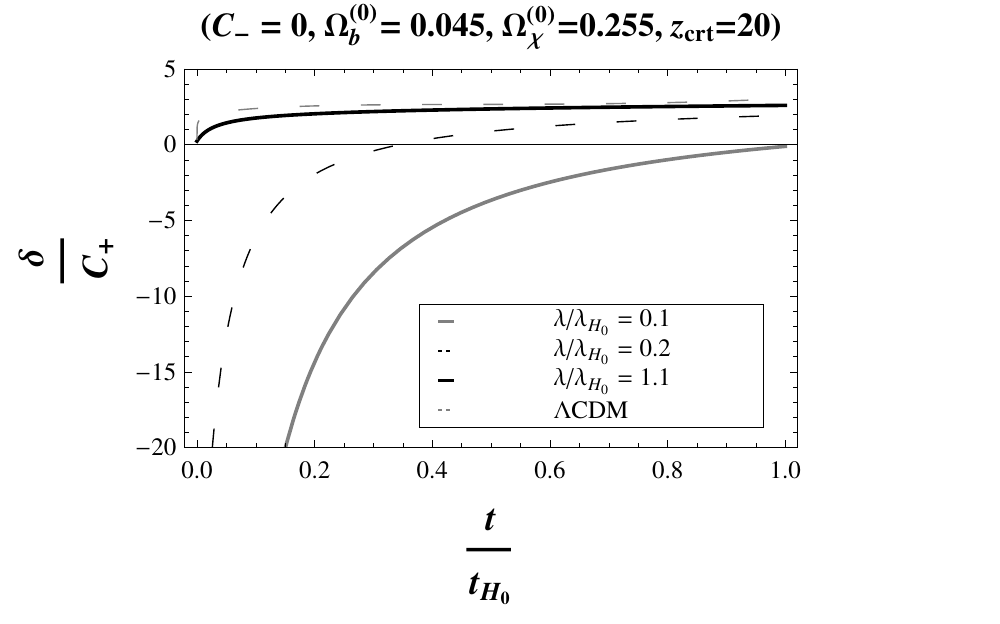}
   \includegraphics[width=0.55\textwidth]{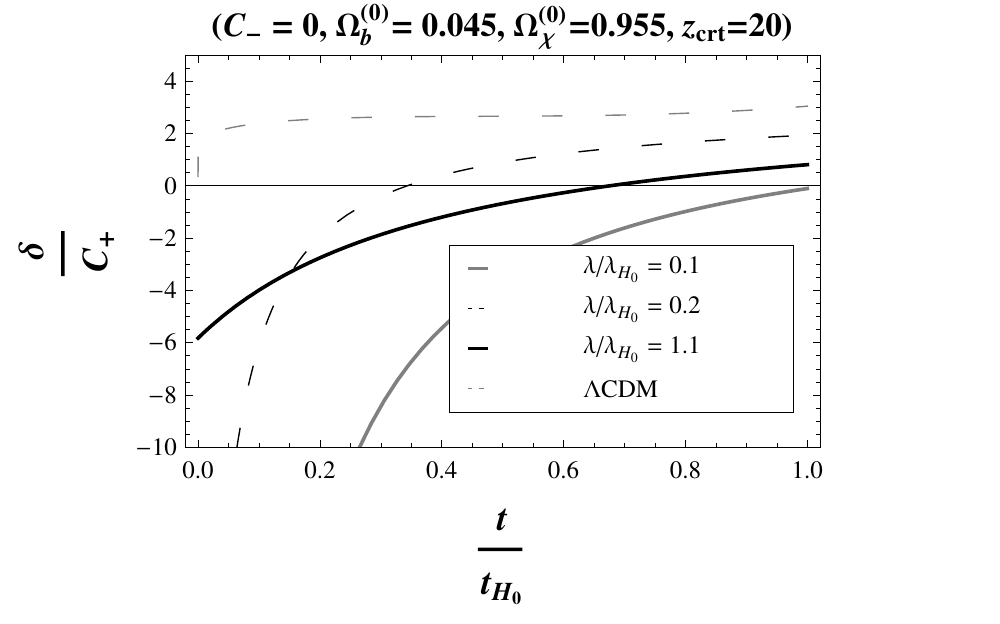} 
   \\
   \includegraphics[width=0.55\textwidth]{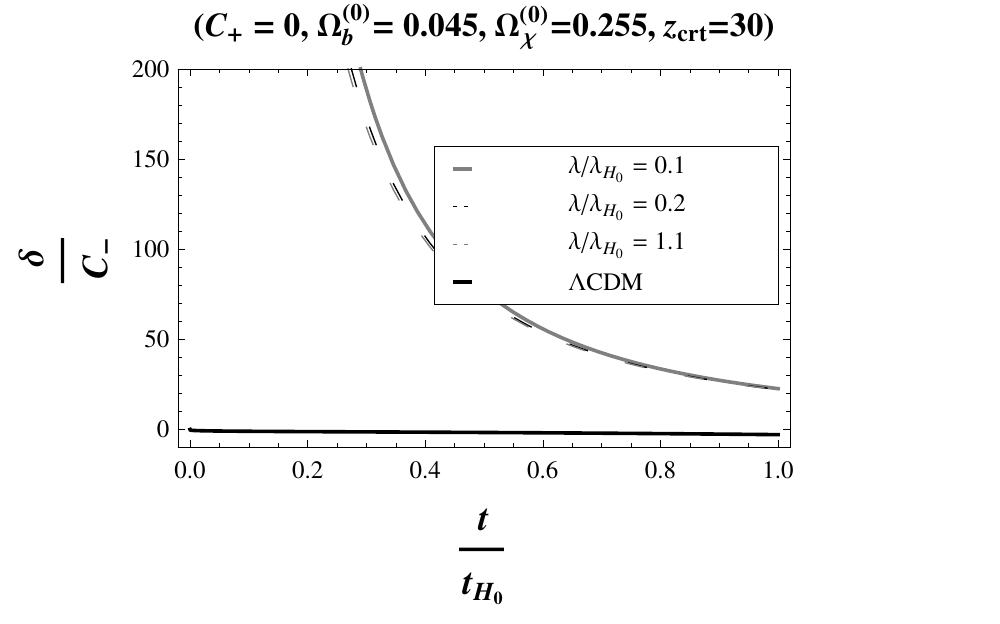}
   \includegraphics[width=0.55\textwidth]{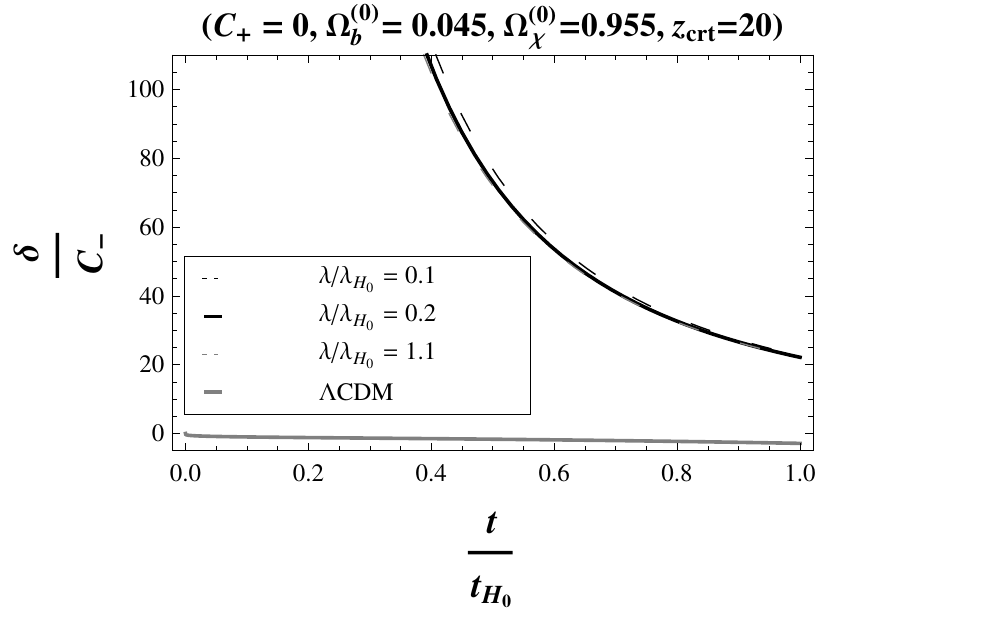}
   \caption{Growing and decaying modes of the matter density contrast as a function of time for various wavelengths for the BEC model and $\Lambda$CDM, where in the BEC model the Universe is filled with baryons and DM, where $\lambda_{H_0} = cH_0^{-1}$ is the Hubble wavelength. In the legends $\Lambda$CDM stands for the concordance model, where $\Omega_b^{0}=0.045$, $\Omega_{dm}^{0}=0.255$ and $\Omega_{\Lambda}=0.7$, and we used in this case $\lambda/\lambda_{H_0}=0.1$.}
   \label{fig.contrastCDM1}
\end{figure}

\begin{figure}[tbp]
   \includegraphics[width=0.55\textwidth]{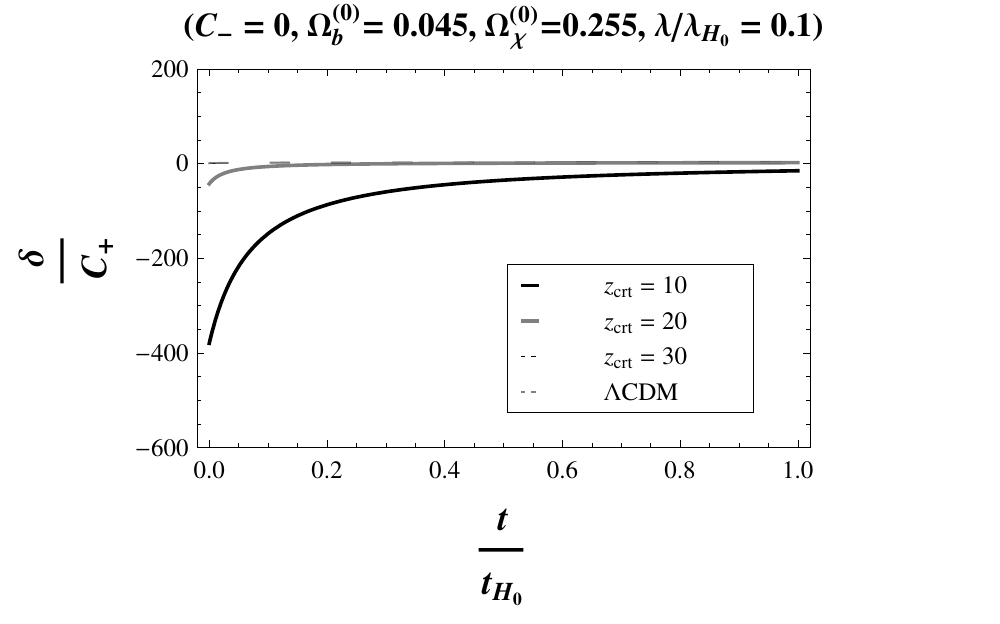}
   \includegraphics[width=0.55\textwidth]{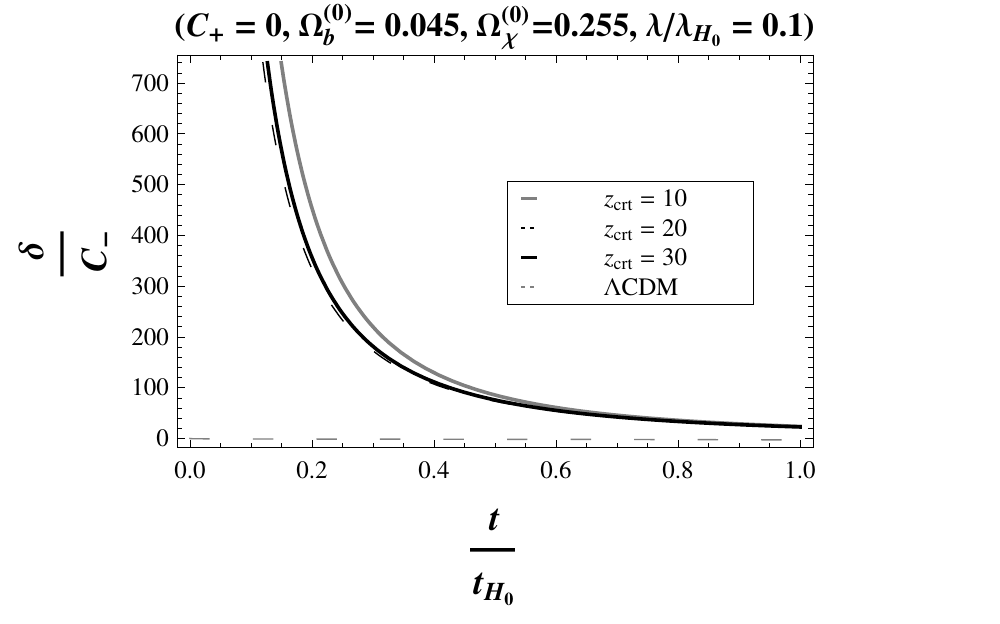}
   \caption{Growing and decaying modes of the matter density contrast as a function of time for various critical redshifts for the BEC model and $\Lambda$CDM, where in the BEC model the Universe is filled with baryons and DM, where $\lambda_{H_0} = cH_0^{-1}$ is the Hubble wavelength. In the legends $\Lambda$CDM stands for the concordance model, where $\Omega_b^{0}=0.045$, $\Omega_{dm}^{0}=0.255$ and $\Omega_{\Lambda}=0.7$, and we used in this case $\lambda/\lambda_{H_0}=0.1$.}
   \label{fig.contrastCDM2}
\end{figure}

Applying the solution (\ref{eq.solescalar1a}) to the equation (\ref{eq.conspert3}) it is straightforward to find
\begin{equation}
  \delta = 3\Phi(R)+\frac{k^2}{a_0^2H_0^2\Omega_{m}}\left[C_{+}\frac{R^{(n_{+}-2)}}{n_{+}(n_{+}-2)}+C_{-}\frac{R^{(n_{-}-2)}}{n_{-}(n_{-}-2)}\right] \quad .
\end{equation}
The growing and decaying modes of the matter density contrast as a function of time for the BEC model with several different values for the model parameters and also for the concordance model ($\Lambda$CDM, with $\Omega_b^{0}=0.045$, $\Omega_{dm}^{0}=0.255$ and $\Omega_{\Lambda}=0.7$) are shown in Figures (\ref{fig.contrastCDM1}) and (\ref{fig.contrastCDM2}).
  
\subsection{Universe filled with baryons, DM and cosmological constant}

In the case of Universe dominated by a cosmological constant with density parameter $\Omega_{\Lambda}$ we find, from equations (\ref{eq.movpertscalar2}), (\ref{eq.omegaaprrox}) and (\ref{eq.vsomaprox}), that
\begin{equation}
    \label{eq.movpertscalar3b}        R^2\frac{d^2\Phi}{dR^2}+\left[4-\frac{1}{2}\left(1-3\frac{\Omega_{\Lambda}R^3}{\Omega_{m}+\Omega_{\Lambda}R^3}\right)\right]R\frac{d\Phi}{dR}+\left[1-\frac{3}{2}\left(1-3\frac{\Omega_{\Lambda}R^3}{\Omega_{m}+\Omega_{\Lambda}R^3}\right)\right]R=0 \quad ,
\end{equation}
and introducing the new variables
\begin{eqnarray}
   x & = & -\frac{\Omega_{\Lambda}R^3}{\Omega_m+\Omega_{\Lambda}R^3} \quad , \\
   \Phi & = & x^{\nu_{\pm}}y(x) \quad ,
\end{eqnarray}
we will have
\begin{equation}
   x(1-x)\frac{d^2y}{dx^2}+\left[c_{\pm}-(a_{\pm}+b_{\pm}+1)x\right]\frac{dy}{dx}-\left(a_{\pm}b_{\pm}\right)y=0 \quad ,
\end{equation}
with
\begin{eqnarray}
   \nu_{\pm} & = & \frac{-5\pm \sqrt{33}}{12} \quad ,\\
   a_{\pm} & = & \frac{1}{4}\pm \frac{\sqrt{33}}{12} \quad , \\
   b_{\pm} & = & a_{\pm} \quad \quad , \\
   c_{\pm} & = & 1\pm \frac{\sqrt{33}}{6} \quad .
\end{eqnarray}

\begin{figure}[tbp]
   \includegraphics[width=0.55\textwidth]{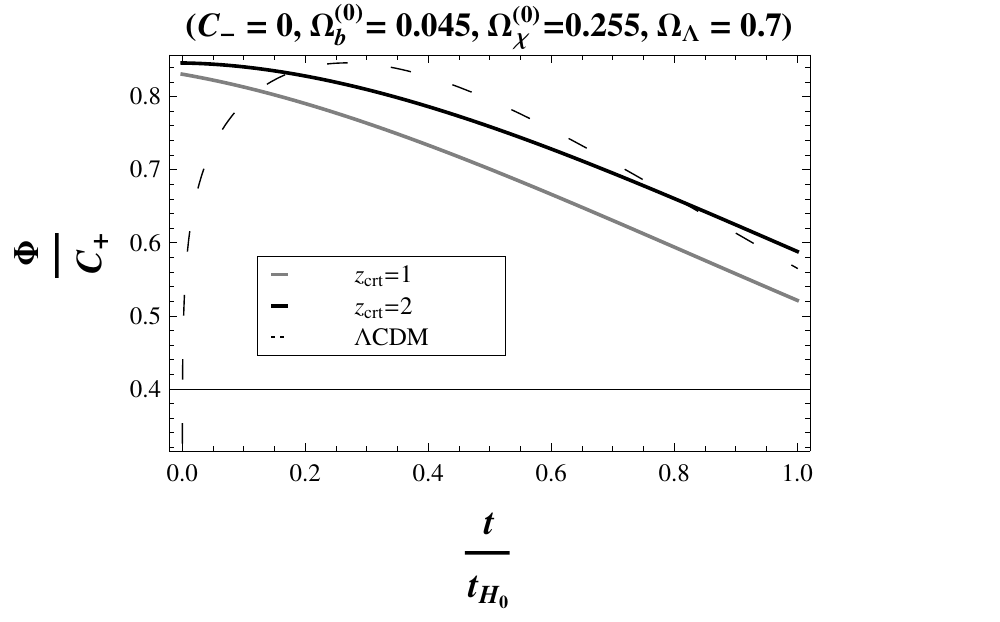}
   \includegraphics[width=0.55\textwidth]{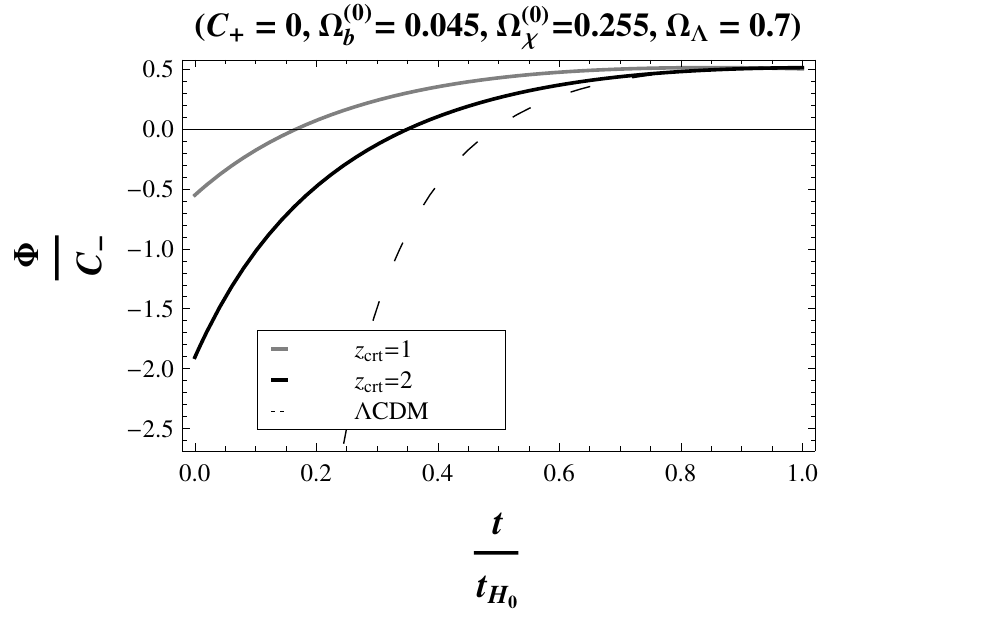}
   \caption{Growing and decaying modes of the potential $\Phi$ as a function of time for the BEC model and $\Lambda$CDM, where in the BEC model the Universe is filled with baryons, DM and cosmological constant. In the legends $\Lambda$CDM stands for the concordance model, where $\Omega_b^{0}=0.045$, $\Omega_{dm}^{0}=0.255$ and $\Omega_{\Lambda}=0.7$.}
   \label{fig.PhiLCDM}
\end{figure}

The solution to the equation (\ref{eq.movpertscalar3b}) is
\begin{equation}
   \label{eq.solPhiLCDM}
   \Phi(R)=C_{+}x^{\nu_{+}}{}_2F_1(a_{+}, b_{+}; c_{+}; x)+C_{-}x^{\nu_{-}}{}_2F_1(a_{-}, b_{-}; c_{-}; x) \quad ,
\end{equation}
where $_2F_1 (a,b;c;x)$ is the hypergeometric function. The growing and decaying modes of the potential $\Phi$ as a function of time for the BEC model with several different values for the model parameters and also for the concordance model ($\Lambda$CDM, with $\Omega_b^{0}=0.045$, $\Omega_{dm}^{0}=0.255$ and $\Omega_{\Lambda}=0.7$) are shown in Figure (\ref{fig.PhiLCDM}).

\begin{figure}[tbp]
   \includegraphics[width=0.55\textwidth]{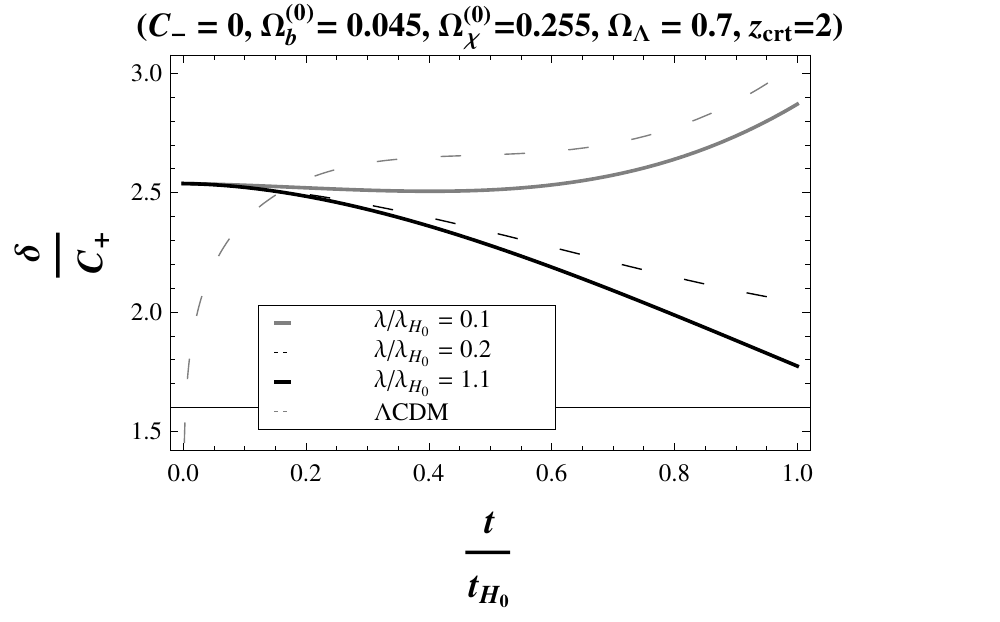}
   \includegraphics[width=0.55\textwidth]{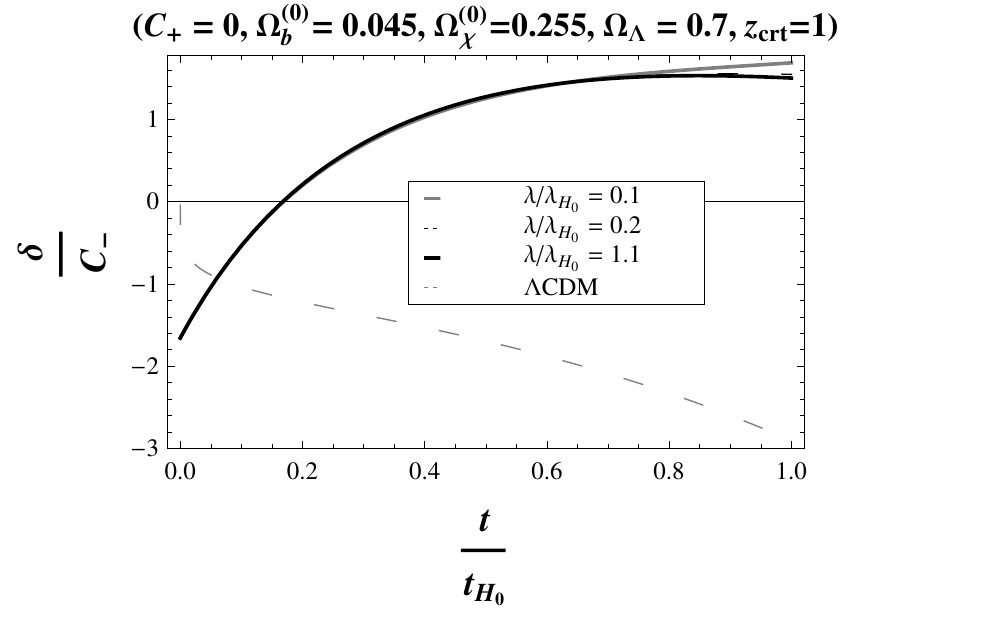} 
   \\
   \includegraphics[width=0.55\textwidth]{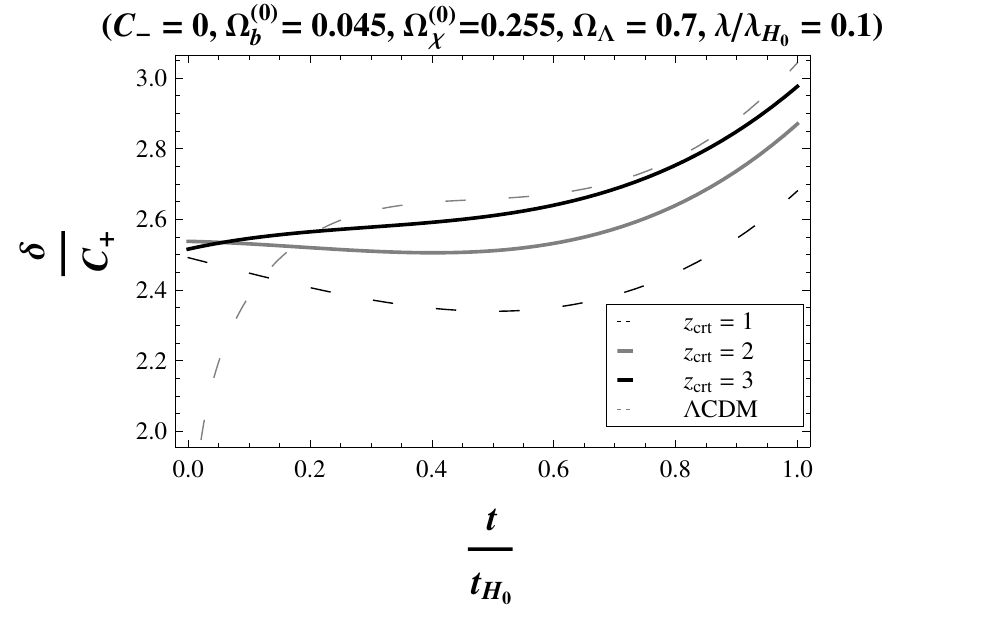}
   \includegraphics[width=0.55\textwidth]{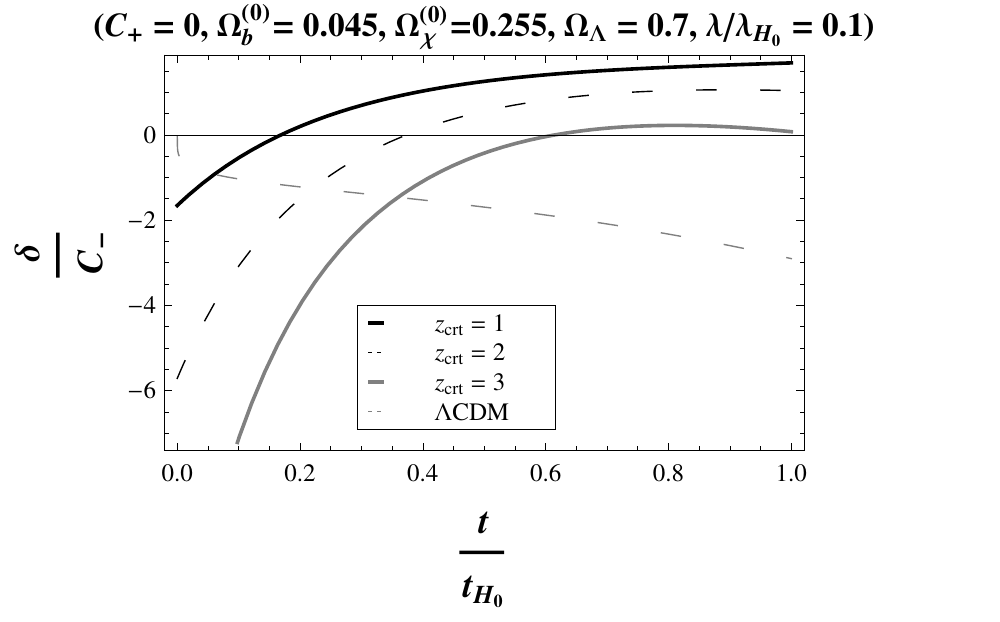}
   \caption{Growing and decaying modes of the matter density contrast as a function of time for various critical redshifts for the BEC model and $\Lambda$CDM, where in the BEC model the Universe is filled with baryons, DM cosmological constant where $\lambda_{H_0} = cH_0^{-1}$ is the Hubble wavelength. In the legends $\Lambda$CDM stands for the concordance model, where $\Omega_b^{0}=0.045$, $\Omega_{dm}^{0}=0.255$ and $\Omega_{\Lambda}=0.7$, and we used in this case $\lambda/\lambda_{H_0}=0.1$}
   \label{fig.contrastLCDM}
\end{figure}

In this case we can solve equation (\ref{eq.conspert3}) numericaly with help of expression (\ref{eq.solPhiLCDM}) to find the density contrast. The growing and decaying modes of the matter density contrast as a function of time are shown in Figure (\ref{fig.contrastLCDM}).

\section{Cosmological gravitational waves}
\label{sec:gw}

In this section we study the gravitational waves generated by the expansion of the Universe during the phase transition of normal DM into BEC DM. To do so it is necessary to perturb to first order the metric tensor and we write
\begin{equation}
   \label{eq.metricpert}
   ds^2=c^2dt^2 - \left[a^2(t)\delta_{ij}+h_{ij}(t,\vec{x})\right]dx^idx^j \quad ,
\end{equation}
where $h_{ij}(t,\vec{x})$ are the metric perturbations and we adopt the transverse and traceless (TT) gauge \cite{muka, maggiore, freitas}, i.e., $h_{i}^{~i}=0$ and $h_{i~,j}^{~j}=0$. Applying the metric (\ref{eq.metricpert}) to the Einstein's field equations (\ref{eq.einstein}) keeping only the first order terms \cite{muka} and with help of the equations of motion (\ref{eq.friedmann}) we find the dynamics equations of the cosmological gravitational waves
\begin{equation}
   \label{eq.onda1}
   \ddot{h}_{ij}-\frac{\dot{a}}{a}\dot{h}_{ij}+\left(-\frac{\nabla^2}{a^2}-\frac{\ddot{a}}{a}\right)h_{ij}=0 \quad ,
\end{equation}
where $\nabla^2$ is the Laplace operator. We introduce the expansion
\begin{equation}
   \label{eq.transf}
   h_{ij}(t,\vec{x})=\sum_{s=\otimes,\oplus}{\int{\frac{d^3k}{(2\pi)^{3/2}}e^{-\dot{\imath}\vec{k}\cdot \vec{x}}h^{(s)}(t,\vec{k})\epsilon^{(s)}_{ij}(\vec{k})}} \quad ,
\end{equation}
where $s=\otimes,\oplus$ are the two polarization states of the gravitational waves, $\vec{k}$ is the wave-number and $\epsilon^{(s)}_{ij}(\vec{k})$ is the polarization tensor, where $\epsilon^{(s)}_{ij}(\vec{k})\epsilon^{(s')ij}(\vec{k'})=2\delta_{ss'}\delta^{3}(\vec{k}-\vec{k'})$ and $h^{(\otimes)}=h^{(\oplus)}=h(t,\vec{k})$. Writing the wave equation (\ref{eq.onda1}) as a function of the scale factor $a$, using the (\ref{eq.transf}) and the background equations (\ref{eq.friedmann}) we find
\begin{equation}
   \label{eq.onda2}
   \frac{d^2h}{da^2}-\frac{3}{2a}\left(1+\omega\right)\frac{dh}{da}+\left[\frac{3k^2}{8\pi G \rho a^4}+\frac{1}{2a^2}\left(1+3\omega\right)\right]h=0 \quad ,
\end{equation}
with $p$ and $\rho$ being the total pressure and density respectively.

With the solution of the wave equation (\ref{eq.onda2}) we can calculate de gravitational waves energy density, that is related to the perturbation $h_{ij}(t,\vec{x})$ by \cite{maggiore}
\begin{equation}
   \rho_{gw}=\frac{1}{32\pi G}\left\langle \dot{h}_{ij}\dot{h}^{ij}\right\rangle \quad ,
\end{equation}
where $\left\langle \cdots \right\rangle$ means spatial average over many reduced wavelengths. The logarithmic energy density is \cite{maggiore}
\begin{equation}
   \Omega_{gw}=\frac{8\pi G}{3H_0^2}\frac{d \rho_{gw}}{d \ln{k}} \quad ,
\end{equation}
and the spectral density $S_h(k)$ \cite{maggiore} is related to the logarithmic energy density by
\begin{equation}
   \Omega_{gw}=\frac{4\pi^2}{3H_0^2}k^3S_h(k) \quad .
\end{equation}

\subsection{Universe filled with baryons and DM}

In order to find a solution to the wave equation (\ref{eq.onda2}) we first consider a Universe filled with baryons, which are pressureless, and DM in phase transition. Since $\sigma^2<<1$  and, in this case, $\Omega_{\Lambda}=0$, we can make the following approximations
\begin{eqnarray}
   \label{eq.aproxCDM}
   \omega \approx 0 \quad , \\
   r \approx n_\chi \quad .
\end{eqnarray}
And the wave equation (\ref{eq.onda2}) will be written as
\begin{equation}
   R^2\frac{d^2h}{dR^2}-\frac{3}{2}R\frac{dh}{dR}+\left(\frac{k^2}{a_0^2H_0^2\Omega_m}R+\frac{1}{2}\right)h=0 \quad ,
\end{equation}
where $R=a/a_0$ and
\begin{equation}
   \Omega_m=\Omega_b^{(0)}+\Omega_\chi^{(0)}(1+z_{crt})^{-3} \quad .
\end{equation}

\begin{figure}[tbp]
   \includegraphics[width=0.55\textwidth]{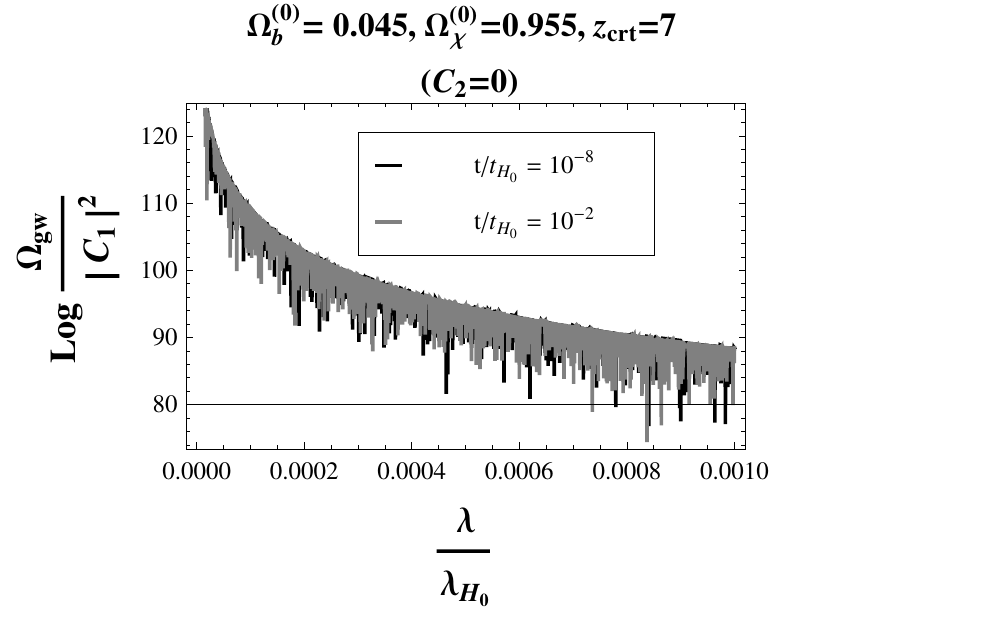}
   \includegraphics[width=0.55\textwidth]{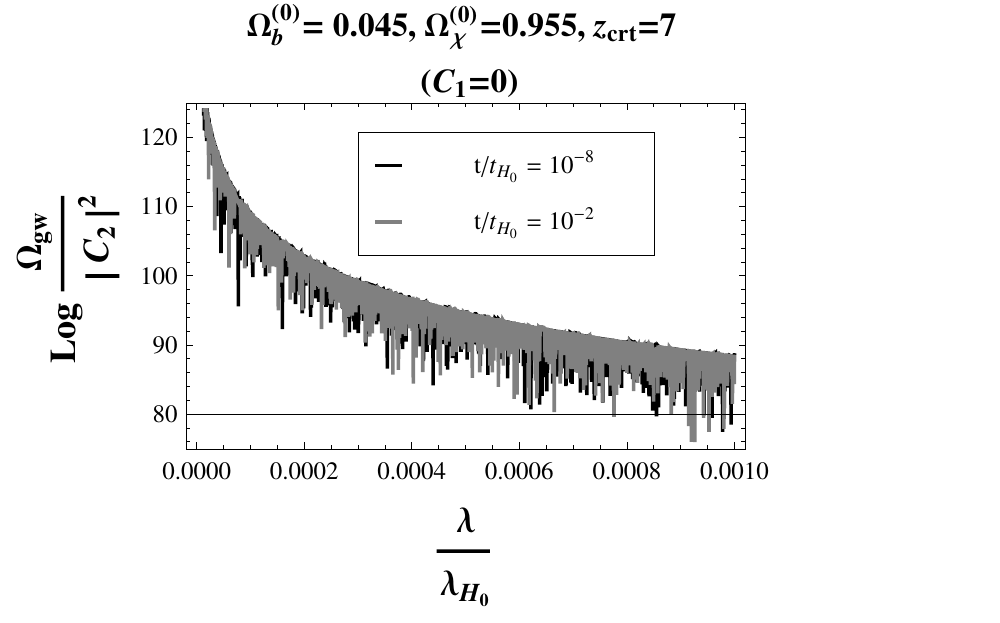} 
   \\
   \includegraphics[width=0.55\textwidth]{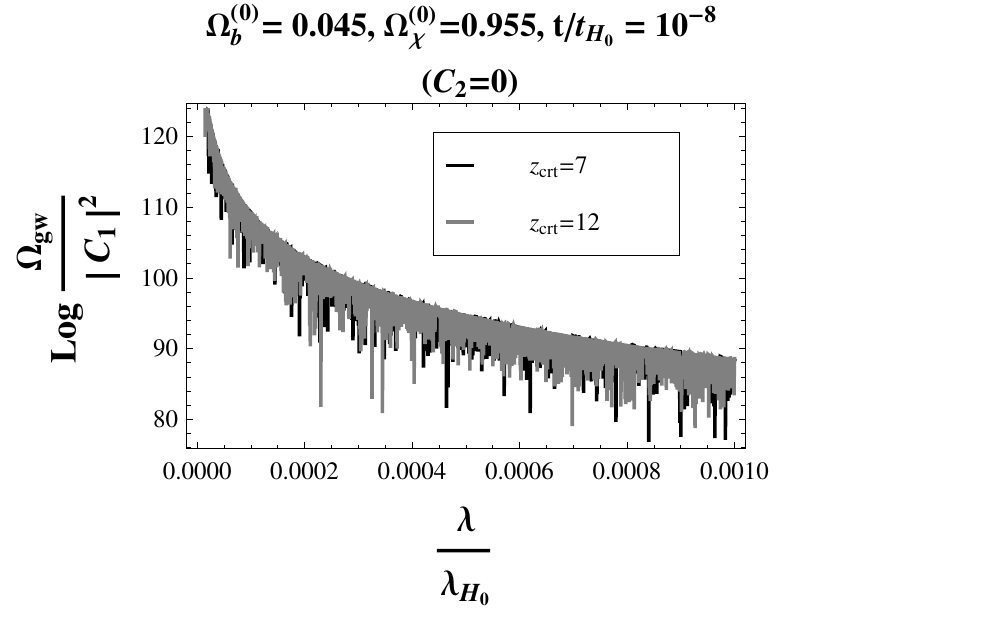}
   \includegraphics[width=0.55\textwidth]{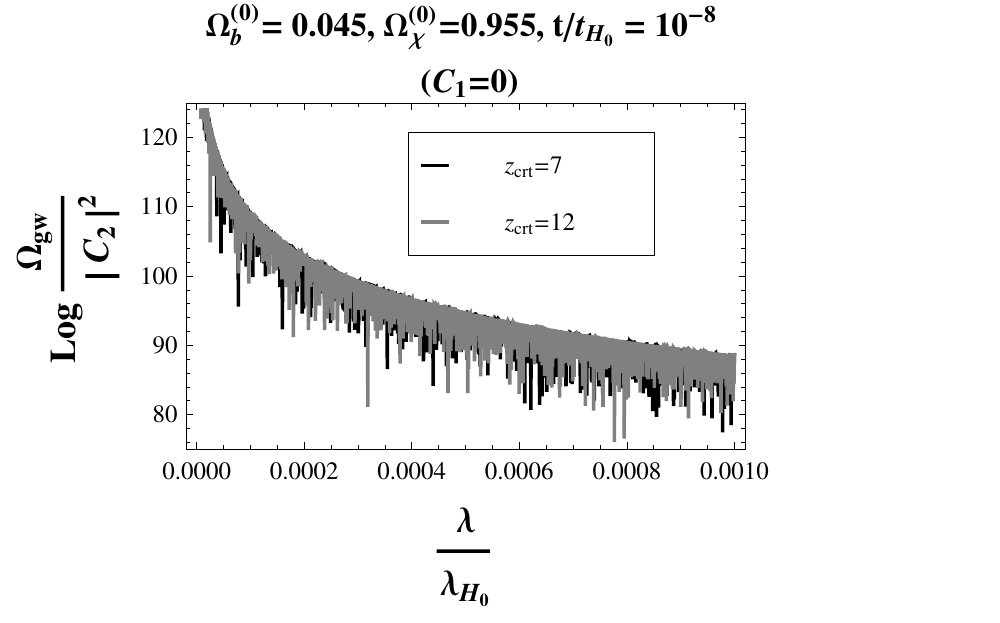}
   \\ 
   \includegraphics[width=0.55\textwidth]{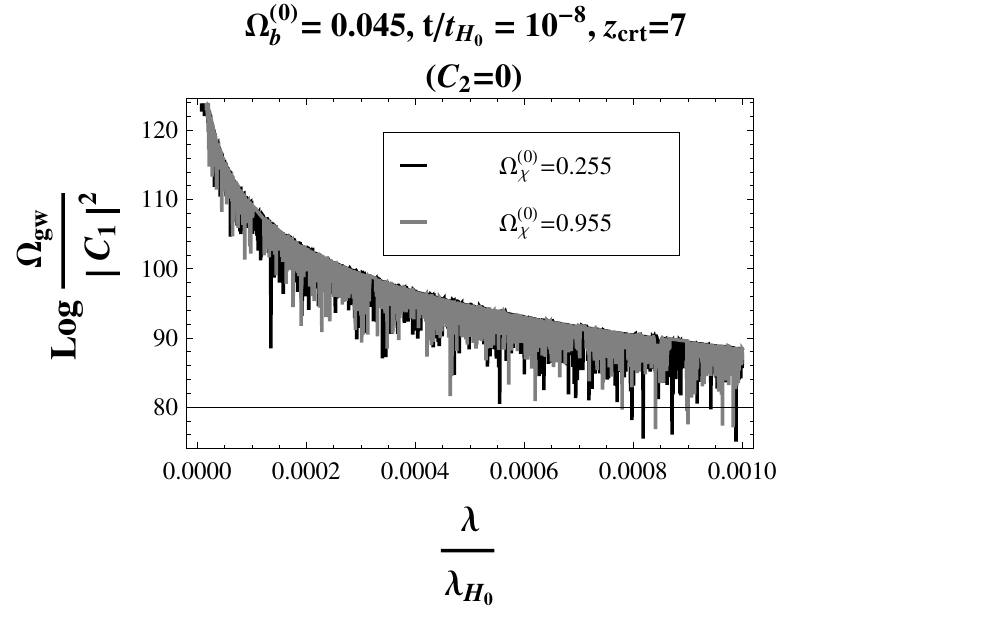}
   \includegraphics[width=0.55\textwidth]{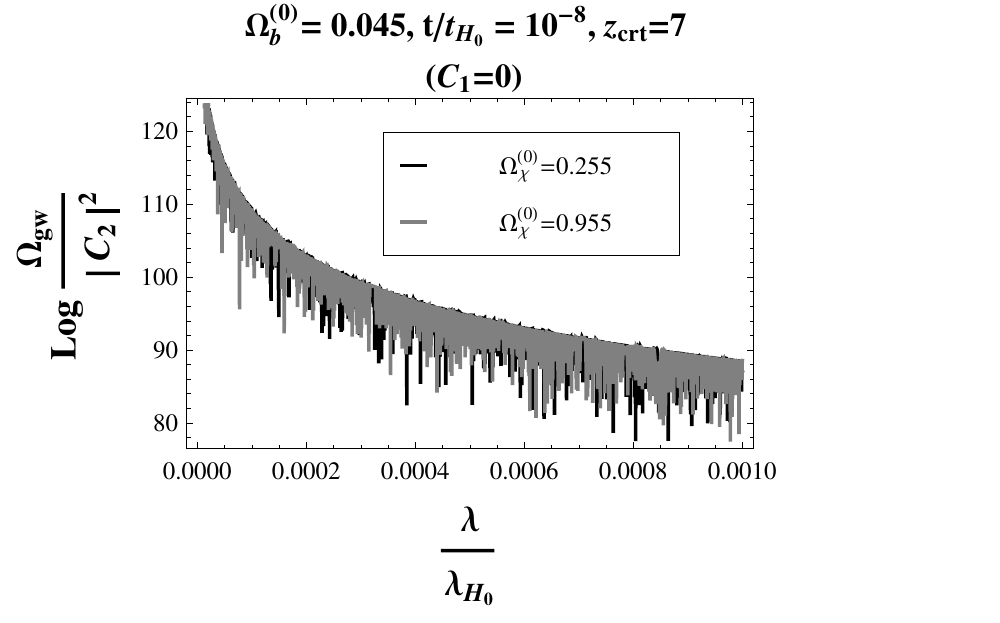}
   \caption{Gravitational waves energy density as a fuction of the wavelength for fixed time in the case of a Universe filled with baryons and DM during the phase transition, where $\lambda_{H_0} = cH_0^{-1}$ is the Hubble wavelength.}
   \label{fig.DenergiaCDM}
\end{figure}

To solve the above differential equation for the gravitational waves it is necessary to introduce the variables
\begin{eqnarray}
   q^2 & \equiv & \frac{k^2}{a^2_0H_0^2\Omega_m} \quad , \\
   x^2 & \equiv & 4Rq^2 \quad , \\
   h & \equiv & x^{5/2}y(x) \quad ,
\end{eqnarray}
and we obtain the Bessel differential equation
\begin{equation}
   x^2y''(x)+xy'(x)+(x^2-17/4)y(x)=0 \quad ,
\end{equation}
which gives the solution
\begin{equation}
   h(2q\sqrt{R})=\left(2q\sqrt{R}\right)^{5/2}\left(C_1J_\nu(2q\sqrt{R})+C_2J_{-\nu}(2q\sqrt{R})\right) \quad ,
\end{equation}
where $J_\nu(x)$ is the Bessel function of first kind with order $\nu = \sqrt{17}/2$ and $C_1$ and $C_2$ are integration constants. In Figure (\ref{fig.DenergiaCDM}) we show the logarithmic energy density for several values of the parameters for a Universe filled with baryons and CDM during the transition phase as a function of the ratio between the gravitational waves wavelength and the Hubble wavelength, defined as $\lambda_{H_0}=cH_0^{-1}$.

\subsection{Universe filled with baryons, DM and cosmological constant}

In this section we will consider a Universe filled with baryons, DM in phase transition and a cosmological constant. In the equation of state parameter (\ref{eq.omega}) the first term can be discharged, since $\sigma^2$ is very small and we can make the approximation
\begin{equation}
   \label{eq.aproxLCDM}
   \omega \approx -\frac{\Omega_{\Lambda}R^3}{\Omega_m+\Omega_{\Lambda}R^3} \quad ,
\end{equation} 
where
\begin{equation}
   \Omega_m=\Omega_b^{0}+\Omega_{\chi}^{(0)}(1+z_{crt})^{-3} \quad ,
\end{equation}
and the wave equation (\ref{eq.onda2}) is written as
\begin{equation}
   \label{eq.onda3} R^2\frac{d^2h}{dR^2}-\frac{3}{2}R\left(\frac{\Omega_m}{\Omega_m+\Omega_{\Lambda}R^3}\right)\frac{dh}{dR}+\left(\frac{q^2R}{\Omega_m+\Omega_{\Lambda}R^3}+\frac{1}{2}\left(1-3\frac{\Omega_{\Lambda}R^3}{\Omega_m+\Omega_{\Lambda}R^3}\right)\right)h=0 \quad ,
\end{equation}
with
\begin{equation}
   q^2=\frac{k^2}{H^2_0a^2_0} \quad .
\end{equation}

For the case of perturbations with wavelength bigger than the Hubble length we approximate $q \approx 0$ to find
\begin{equation}
   \label{eq.onda4a} R^2\frac{d^2h}{dR^2}-\frac{3}{2}R\left(\frac{\Omega_m}{\Omega_m+\Omega_{\Lambda}R^3}\right)\frac{dh}{dR}+\frac{1}{2}\left(1-3\frac{\Omega_{\Lambda}R^3}{\Omega_m+\Omega_{\Lambda}R^3}\right)h=0 \quad ,
\end{equation}
and we introduce the variables
\begin{eqnarray}
   \label{eq.transf1}
   x = -\frac{\Omega_{\Lambda}}{\Omega_m}R^{3} \quad , \\
   \label{eq.transf2}
   h(x) = x^{\nu_{\pm}}y(x) \quad ,
\end{eqnarray}
where $\nu_{\pm}=5/12\pm\sqrt{17}/12$ and we find
\begin{equation}
   \label{eq.onda5}
   x(1-x)\frac{d^2y}{dx^2}+[c_{\pm}-(a_{\pm}+b_{\pm}+1)x]\frac{dy}{dx}-a_{\pm}b_{\pm}y=0 \quad ,
\end{equation}
with 
\begin{eqnarray}
   a_{\pm} & = & \frac{(3-2\sqrt{5}\pm\sqrt{17})}{12} \quad , \\
   b_{\pm} & = & \frac{(3+2\sqrt{5}\pm\sqrt{17})}{12}  \quad , \\
   c_{\pm} & = & 1\pm \frac{\sqrt{17}}{6} \quad .
\end{eqnarray}

The solution is found to be
\begin{equation}
   h(R)=C_{+}x^{\nu_{+}}{}_2F_1(a_{+}, b_{+}; c_{+}; x)+C_{-}x^{\nu_{-}}{}_2F_1(a_{-}, b_{-}; c_{-}; x) \quad ,
\end{equation}
where ${}_2F_1(a, b; c; x)$ is the hypergeometric function.

\begin{figure}[tbp]
   \includegraphics[width=0.5\textwidth]{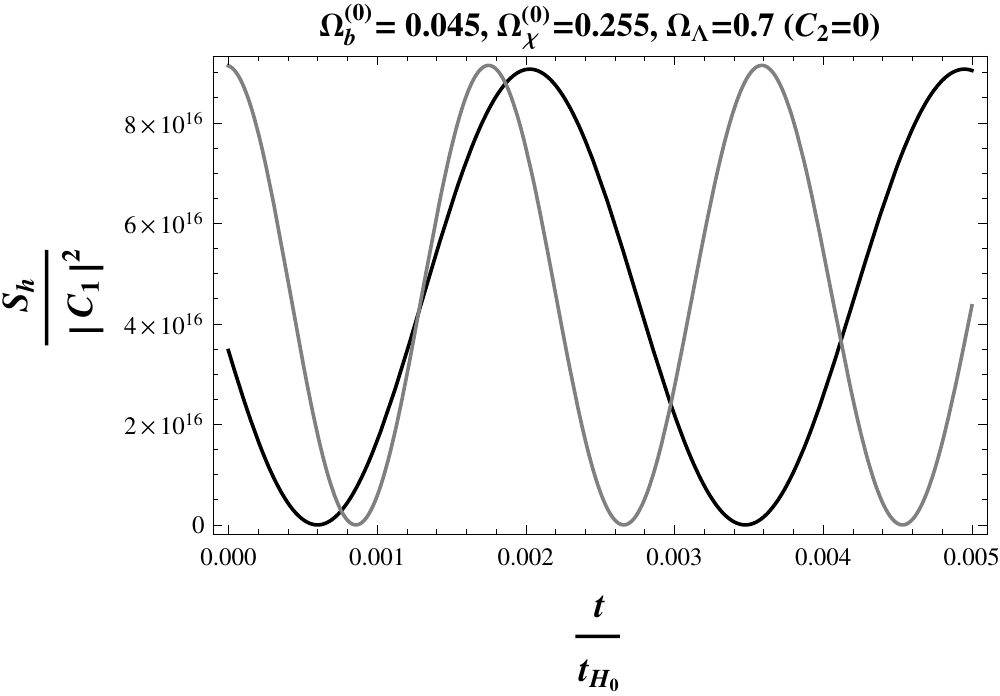}
   \includegraphics[width=0.5\textwidth]{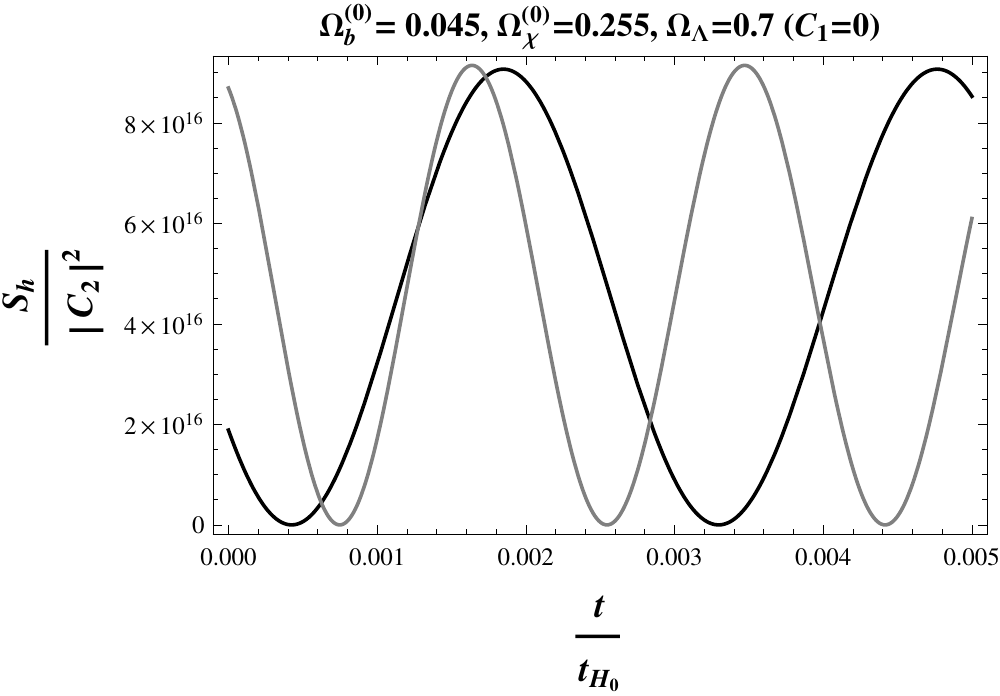} 
   \caption{Gravitational waves spectral density as a fuction of time in the case of a Universe filled with baryons and DM during the phase transition. The black curve shows the case where $z_{crt}=7$ and in the gray one $z_{crt}=12$.}
   \label{fig.SpecLCDM}
\end{figure}

For small wavelenghts, the equation (\ref{eq.onda3}) will be
\begin{equation}
   \label{eq.onda4b} R^2\frac{d^2h}{dR^2}-\frac{3}{2}R\left(\frac{\Omega_m}{\Omega_m+\Omega_{\Lambda}R^3}\right)\frac{dh}{dR}+\frac{q^2R}{\Omega_m+\Omega_{\Lambda}R^3}h=0 \quad ,
\end{equation}
which can only be solved numericaly.
In Figure (\ref{fig.SpecLCDM}) we show the spectral density, that in this case does not depend on the wavelength, as a function of time in a Universe filled with baryons CDM and cosmological constant during the transition.

\section{Conclusions}
\label{sec:conc}

In this letter we considered a Universe where initially the DM is nearly pressureless and after a phase transition the DM is all in form of BEC DM, which has a polytropic equation of state. During the transition phase both non-condensated and condensated DM coexist and the time evolution of the Universe expansion is changed. We studied the scalar and tensorial perturbations generated by the background expansion during the phase transition in two cases: with and without cosmological constant.

Firstly, we verify from Figure \ref{fig.hubbcrit} and from equations (\ref{eq.hubblepar1}, \ref{eq.hubblepar2}) that the ratio between the Hubble parameter at the beginning of the condensation process and the Hubble parameter today, with the condition $H_{crt}/H_0 > 1$, is satisfied for $z_{crt} > 3.5$, $z_{crt} > 3.75$ and $z_{crt} > 0$, what was expected. We can see that the case where $\Lambda$ is present is favored in the sense that the ratio of the Hubble parameters is bigger than the case where the cosmologcial constant is absent. This indicates that in a Universe whose material content is baryonic matter, BEC DM and cosmological constant the condensation can begin later when compared with the model without $\Lambda$.

When we analyze the scalar potential $\Phi$ we see that in the case without cosmological constant, Figures \ref{fig.PhiCDM}, there is a growing and a decaying mode. Both modes are more sensitivy to changes in the amount of DM than to changes in the critical redshift $z_{crt}$. When compared with the standard $\Lambda$CDM model we see that, in the case of the growing mode, the BEC model can poorly reproduce the behavior of the concordance model. But for the decaying model it is evident that a small amount of BEC DM and small $z_{crt}$ are favored.

In the Universe with cosmological constant, Figures \ref{fig.PhiLCDM}, we see two decaying modes and smaller the critical redshfit $z_{crt}$ smaller the potential $\Phi$. When compared with the concordance model ($\Lambda$CDM) big values for the critical redshift $z_{crt}$ are prefered, although for late times the models seem to be indistinguishable as seen in the case of de decaying modes.

In Figures \ref{fig.contrastCDM1} and Figure \ref{fig.contrastCDM2} we see that the density contrast amplitude, in a Universe without cosmological constant, is sensitivy to the amount of DM and bigger the perturbation wavelength bigger the perturbation in the case of the growing modes and smaller in the case of the decaying modes. The same behavior can be seen when the critical redshfit is varied. Once again, when compared with the concordance model this model can poorly reproduce his behavior, especially for earlier times.

For a Universe with cosmological constant the contrast amplitude in Figure \ref{fig.contrastLCDM} can not be distinguished when we vary the perturbations wavelength in the case of the decaying modes. For the growing modes the amplitude initially decays but after some time interval it grows again. The decaying period of the growing modes depends on the wavelength. The bigger the wavelength bigger is the decaying time interval. The same behavior is reproduced when we vary the critical redshift $z_{crt}$. And it is evident from the Figure \ref{fig.contrastLCDM} that big values of the critical redshift $z_{crt}$ are prefered to reproduce the concordance model.

It is very important to note that critical redshift $z_{crt}$ is the most relevant parameter in this analysis. This parameter allows us to distinguish between our model and the concordance model and contains some important informations about the BEC DM and the condensation process. As can be seen from equation (\ref{eq:zcrit}) for a fixed scattering length parameter $l_s$, which has a typical value from $10^6~\textrm{fm}$ to $10^{10}~\textrm{fm}$, big values for the particle mass are favored in order to have big values of the critical redshift. Considering this range of values for the scattering length parameter and the maximum value of $1.87~\textrm{eV}$ for the DM particle mass it is possible to have a maximum critical redshift $z_{crt}$ between $17$ and $400$, which are bigger than the examples in the figures. In \cite{Vel} it was shown that even big variations in the particle mass do not allow to discriminate between the BEC model and the concordance model.

In the case of tensorial perturbations in a Universe without cosmological constant we see in Figures \ref{fig.DenergiaCDM} that the gravitational waves energy density changes little when we vary the time of beginning of the transition ($z_{crt}$) and even the amount of DM in the Universe. Figures \ref{fig.SpecLCDM} shows little changes in the spectral density in a Universe with cosmological constant when we change the beginning of the transition.

So that we analyzed in this work a particular phase of evolution of the Universe, where non-condensated and condensated DM coexist, it is difficult to make comparisons with other cosmological models and their parameters (like the timescale $H_0 \propto 1/t_0$, the density parameters $\Omega_i$, the equation of state parameter $\omega = p/\rho$, the scalar spectral index $n$, the Tensor-to-scalar ratio $r$, etc), mainly with the simplest of them the $\Lambda$CDM model. We need to evolve the Universe from a primordial inflation phase and calculate the specific cosmological parameters and thus compare them with the values of the parameters obtained in other cosmological models.

Though, in general, the Bose-Einstein condensate dark matter model can be an interesting theory to explain the nature of dark matter in the Universe. A more detailed investigation of the model discussed herein, including the generalization of the equation of state of BEC dark mater model, is being made as well as the bayesian analysis to determine the most probable values of the various parameters of the theory. Perchance only when we compare the model with the data it will be possible to determine the value of $z_{crt}$ and conclude if it is possible to dintinguish between BEC DM and the ordinary CDM. This comparison is not straightforward and is beyond the scope of this work. The BEC DM model still has to pass by several observational tests to prove to be a viable cosmological model. Its importance grows as weakly interacting massive particles, or WIMPs, are searched in the Large Hadron Collider or other ground-based experiments.

\acknowledgments
This work has received partial financial supporting from CNPq (Brazil) and CAPES (Brazil). We thanks the referee for the helpful comments.

\end{document}